\documentclass{emulateapj}
\usepackage{apjfonts}
\newcommand{\figexpand}{\epsscale{1.15}}
\newcommand{\tableclear}{}
\newcommand{\tableset}{deluxetable*}
\newcommand{\plotter}{\plotone}

\newcommand{\etal}{et al.}
\newcommand{\mbh}{M_{\rm BH}}
\newcommand{\mstar}{M_{\ast}}
\newcommand{\mdyn}{M_{\rm dyn}}
\newcommand{\mhalo}{M_{\rm halo}}
\newcommand{\re}{R_{e}}
\newcommand{\vvir}{V_{\rm vir}}
\newcommand{\fgas}{f_{\rm gas}}
\newcommand{\sersic}{n_{s}}
\newcommand{\msun}{M_{\sun}}
\newcommand{\tH}{t_{\rm H}}
\newcommand{\mdotstar}{\dot{M}_{\ast}}
\newcommand{\paperone}{Paper \textrm{I}}

\shorttitle{Modeling the Black Hole Fundamental Plane}
\shortauthors{Hopkins \etal}
\slugcomment{Accepted to ApJ, May 2007}
\begin{document}

\title{A Theoretical Interpretation of the Black Hole Fundamental Plane}
\author{Philip F.\ Hopkins\altaffilmark{1}, 
Lars Hernquist\altaffilmark{1}, 
Thomas J. Cox\altaffilmark{1}, 
Brant Robertson\altaffilmark{2}, \&\
Elisabeth Krause\altaffilmark{3}
}
\altaffiltext{1}{Harvard-Smithsonian Center for Astrophysics, 60 Garden Street, Cambridge, MA 02138}
\altaffiltext{2}{Kavli Institute for Cosmological Physics, The University of Chicago, 
5460 S.\ Ellis Ave., Chicago, IL 60637}
\altaffiltext{3}{Department of Physics and Astronomy, Universit\"{a}t Bonn, 53121 Bonn, Germany}

\begin{abstract}

We examine the origin and evolution of correlations between properties
of supermassive black holes (BHs) and their host galaxies using
simulations of major galaxy mergers, including the effects of gas
dissipation, cooling, star formation, and BH accretion and feedback.
We demonstrate that the simulations predict the existence of a BH
``fundamental plane'' (BHFP), of the form
$\mbh\propto\sigma^{3.0\pm0.3}\,\re^{0.43\pm0.19}$ or
$\mbh\propto\mstar^{0.54\pm0.17}\,\sigma^{2.2\pm0.5}$, similar to
relations found observationally.  The simulations indicate that the
BHFP can be understood roughly as a tilted intrinsic correlation
between BH mass and spheroid binding energy, or the condition for
feedback coupling to power a pressure-driven outflow.  While changes
in halo circular velocity, merger orbital parameters, progenitor disk
redshifts and gas fractions, ISM gas pressurization, and other
parameters can drive changes in e.g.\ $\sigma$ at fixed $\mstar$, and
therefore changes in the $\mbh-\sigma$ or $\mbh-\mstar$ relations, the
BHFP is robust. Given the empirical trend of decreasing $\re$ for a
given $\mstar$ at high redshift (i.e.\ increasingly deep potential
wells), the BHFP predicts that BHs will be more massive at fixed
$\mstar$, in good agreement with recent observations. This evolution
in the structural properties of merger remnants, to smaller $\re$ and
larger $\sigma$ (and therefore larger $\mbh$, conserving the BHFP) at
a given $\mstar$, is driven by the fact that disks (merger
progenitors) have characteristically larger gas fractions at high
redshifts. Adopting the observed evolution of disk gas fractions with
redshift, our simulations predict the observed trends in both
$\re(\mstar)$ and $\mbh(\mstar)$. The existence of this BHFP also has
important implications for the masses of the very largest black holes,
and immediately resolves several apparent conflicts between the BH
masses expected and measured for outliers in both the $\mbh-\sigma$
and $\mbh-\mstar$ relations.

\end{abstract}


\section{Introduction}
\label{sec:intro}

Correlations between the masses of supermassive black holes (BHs) in
the centers of galaxies and the properties of their host spheroids
\citep[e.g.,][]{KormendyRichstone95} imply a fundamental bond between
the growth of BHs and galaxy formation.  A variety of such
correlations have been identified, linking BH mass to host luminosity
\citep{KormendyRichstone95}, mass \citep{magorrian}, velocity
dispersion \citep{FM00,Gebhardt00}, concentration or Sersic index
\citep{graham:concentration,graham:sersic}, and binding energy
\citep{aller:mbh.esph}.  However, the connection between these
relationships is obscured by the fact that the properties of the host
spheroids are themselves correlated \citep[see e.g.][for a comparison
of the observed relations]{novak:scatter}.  The lack of a clear
motivation for favoring one correlation over another has led to
considerable debate over the interpretation of systems deviating from
the mean correlation between host properties, and over the
demographics of the most massive BHs
\citep[e.g.,][]{bernardi:magorrian.bias,
lauer:massive.bhs,batcheldor:bcgs,wyithe:log.quadratic.msigma}.

Analytical estimates
\citep[e.g.,][]{silkrees:msigma,burkertsilk:msigma} and studies using
simulations \citep{cox:kinematics,robertson:fp} have demonstrated that
these correlations, in particular the $\mbh-\sigma$
\citep{dimatteo:msigma} and $\mbh-\mstar$
\citep{robertson:msigma.evolution} relations, can be reproduced in
feedback-regulated models of BH growth.  However, determining the
fundamental character and evolution of these correlations with
redshift is critical for informing analytical models
\citep[e.g.,][]{croton:msigma.evolution} and simulations
\citep{dimatteo:msigma,robertson:msigma.evolution,
hopkins:lifetimes.letter} which follow the
co-formation of BHs and bulges, as well as theories which relate the
evolution and statistics of BH formation and quasar activity to galaxy
mergers
\citep[e.g.,][]{hopkins:qso.all,hopkins:bol.qlf,hopkins:merger.lfs,
hopkins:frame1}
and to the remnant spheroid population \citep{hopkins:red.galaxies,
hopkins:frame2}.
Likewise, the significance of observations tracing the buildup of
spheroid populations \citep[e.g.,][]{Cowie96} and associations between
spheroids in formation, mergers, and quasar hosts
\citep{hopkins:transition.mass} depends on understanding the evolution
of BH/host correlations.

Unfortunately, efforts to directly infer these correlations at redshifts $z\gg0$ are
difficult and still limited by the small number of observable hosts. 
Furthermore, without understanding the fundamental nature of the correlations between 
BH and host properties, it is difficult to interpret these observations, as they do 
not all probe the same correlations. 
Consequently, different groups have arrived at
seemingly contradictory conclusions. Velocity dispersion measurements have favored 
both no evolution \citep[][from OIII velocity dispersions]{shields03:msigma.evolution} 
and substantial evolution 
\citep[][from CO dispersions and spectral template fitting]{shields06:msigma.evolution,
woo06:lowz.msigma.evolution}. 
BH clustering measurements \citep{adelbergersteidel:magorrian.evolution,
wyithe:magorrian.clustering,hopkins:clustering,lidz:clustering} 
suggest moderate evolution in the 
ratio of BH to host halo mass at redshifts $z\sim1-3$. Direct host $R$-band luminosity 
measurements \citep{peng:magorrian.evolution} and indirect comparison of 
quasar luminosity and stellar mass densities \citep{merloni:magorrian.evolution} 
or BH and stellar mass functions \citep{hopkins:msigma.limit} similarly 
favor moderate evolution in the ratio of BH to host spheroid stellar mass 
occurring at $z\gtrsim1$, and dynamical masses from CO measurements 
suggest that this evolution may extend to $z\sim6$ \citep{walter04:z6.msigma.evolution}. 
Better understanding the dependence of 
BH mass on host properties can provide both a self-consistent 
paradigm in which to interpret these observations and 
(potentially) a physically and observationally motivated prediction of 
their evolution. 

One possibility is that these different correlations are projections of the 
same ``fundamental plane'' (FP) relating BH mass with two or more spheroid 
properties such as stellar mass, velocity dispersion, or effective radius, in analogy to 
the well-established fundamental plane of spheroids. For the case of spheroids, it is now 
understood that various correlations, including the Faber-Jackson relation 
\citep{fj76} between luminosity (or effectively stellar mass $\mstar$) 
and velocity dispersion 
$\sigma$, the \citet{kormendy77:correlations} relation between effective radius $\re$ and surface 
brightness $I_{e}$, and the size-luminosity or size-mass relations \citep[e.g.,][]{shen:size.mass} 
between $\re$ and $\mstar$, are all projections of a fundamental plane relating 
$\re\propto \sigma^{\alpha}\,I_{e}^{\beta}$ \citep{dressler87:fp,dd87:fp}. 

In their analysis of the relation between BH mass and host luminosity or 
dynamical mass, $\mdyn$, \citet{marconihunt} \citep[see also][]{defrancesco:mbh.mdyn} 
noted that the residuals of 
the $\mbh-\sigma$ relation \citep[effectively $\mbh/\sigma^{4}$;][]{tremaine:msigma} were significantly 
correlated with the effective radii of the systems in their sample. 
A more detailed study by \citet{hopkins:bhfp.obs} (henceforth \paperone) 
found that the observations indeed 
favor a fundamental plane-type relation between $\mbh$ and 
any two of $\re$, $\sigma$, or $\mstar$, at $>3\,\sigma$ confidence. This 
observed, low redshift BHFP has the form 
$\mbh\propto\sigma^{3.0\pm0.3}\,\re^{0.43\pm0.19}$ or
$\mbh\propto\mstar^{0.54\pm0.17}\,\sigma^{2.2\pm0.5}$, with 
the early-type FP providing a tight mapping between $\mstar$, $\re$, and $\sigma$. 
Given the mean correlations between e.g.\ $\re$ and $\mstar$ or $\mstar$ and $\sigma$, 
the previously recognized correlations between BH mass and 
spheroid mass, luminosity, velocity dispersion, and binding energy can all 
naturally be explained as projections of this intrinsic BHFP.
However, these observations remain limited in the range of systems they probe, 
and are restricted to passive, local spheroids. We therefore wish both to 
understand the origin of this BHFP and how it should (or should not) evolve with redshift. 

In this paper, we investigate the nature of the correlation between BH
mass and host properties and the existence of a fundamental plane
relating BH mass and spheroid mass, velocity dispersion, and effective
radius.  In \S~\ref{sec:sims}, we describe a large set of numerical
simulations which we use to study and predict the nature of the
BH-host correlations under a wide variety of conditions, and in
\S~\ref{sec:data} we describe the observational data sets we compile
to study the observed correlations.  In \S~\ref{sec:local} we describe
the correlations determined from both simulations and observations,
and then analyze the correlations between residuals in e.g.\ the
$\mbh-\sigma$ relation and secondary properties such as $\re$ and
$\mstar$, which leads us in \S~\ref{sec:local:FP} to discuss the
fundamental plane relating BH mass and $\sigma$, $\re$, and
$\mstar$. \S~\ref{sec:pred.local} discusses the implications of this
relation for predicting BH masses and demographics, and
\S~\ref{sec:origins} considers the physical origin of the BHFP
relation. In \S~\ref{sec:driving}, we study how various theoretical
quantities or initial conditions drive systems along the BHFP relation
and, as a consequence, drive evolution in the various projections of
the BHFP, and in \S~\ref{sec:evolution} apply this to understand the
observed evolution with redshift in the $\mbh-\mstar$ and
$\mbh-\sigma$ relations. We summarize our conclusions and discuss
future tests of our proposed relations in \S~\ref{sec:discussion}.

Throughout, we adopt a $\Omega_{\rm M}=0.3$, $\Omega_{\Lambda}=0.7$,
$H_{0}=70\,{\rm km\,s^{-1}\,Mpc^{-1}}$ cosmology 
(and correct all observations accordingly), but note this choice has little 
effect on our conclusions.

\section{The Data}
\label{sec:data}

The observational data set with which we compare 
is described in detail in \paperone, but we 
summarize it here.
We consider the sample of local BHs for which masses have been reliably determined 
via either kinematic or maser measurements. Specifically, we adopt the sample of 
38 local systems for which values of $\mbh$, $\sigma$, $\re$, $\mdyn$, 
and bulge luminosities are compiled in 
\citet{marconihunt} and \citet{haringrix} \citep[see also][]{magorrian,
merrittferrarese:msigma,tremaine:msigma}. 
We adopt the dynamical masses from the more detailed Jeans modeling 
in \citet{haringrix}. We estimate the total stellar mass $\mstar$ from the total
$K$-band luminosity given in \citet{marconihunt}, using the $K$-band mass-to-light 
ratios as a function of luminosity from \citet{bell:mfs} (specifically assuming a 
``diet'' Salpeter IMF, although this only affects the absolute normalization of 
the relevant relations). In \paperone\ it is also 
noted that the choice of these mass-to-light 
ratios as opposed to those determined from e.g.\ photometric fitting makes little 
difference. 
We take measurements of the Sersic index $\sersic$ from 
\citet{graham:sersic}. Where possible, we update measurements of $\re$, $\sigma$ and $\sersic$ with 
more recent values from \citet{lauer:centers,lauer:bimodal.profiles,mcdermid:sauron.profiles} and 
from \citet{kormendy:wetvsdry}, which extends the baseline of 
surface brightness measurements allowing more robust estimates of $\sersic$ and $\re$.

Although it should only affect the normalization of the relations
herein, we note that our adopted cosmology is identical to that used
to determine all quoted values in these works.  The concentration
index $R_{30}/R_{50}$ for the observed systems is calculated assuming
a Sersic profile with the best-fit $\sersic$. When we fit the
observations to e.g.\ the mean $\mbh-\sigma$ relation and other
BH-host relations, we consider only the subsample of 27 objects in
\citet{marconihunt} which are deemed to have `secure' BH and bulge
measurements (i.e.\ for which the BH sphere of influence is clearly
resolved, the bulge profile can be well-measured, and maser spots
(where used to measure $\mbh$) are in Keplerian orbits).  Our results
are not qualitatively changed if we consider the entire sample in
these fits, but their statistical significance is somewhat reduced.

\section{Simulations}
\label{sec:sims}

\subsection{Methodology}
\label{sec:sims:methodology}

Our simulations are taken from \citet{robertson:fp}, who utilize a set of several 
hundred simulations of major mergers to study the properties 
of remnants on the early-type galaxy FP. 
The properties of the models
are discussed in detail therein, but we briefly review them here. 
The simulations were performed with the parallel TreeSPH code
{\small GADGET-2} \citep{springel:gadget}, based on a fully conservative
formulation \citep{springel:entropy} of smoothed particle hydrodynamics (SPH),
which conserves energy and entropy simultaneously even when smoothing
lengths evolve adaptively \citep[see, e.g.,][]{hernquist:sph.cautions,oshea:sph.tests}. 
Our simulations account for radiative cooling, heating by
a UV background \citep[as in][]{katz:treesph,dave:lyalpha}, and
incorporate a sub-resolution model of a multiphase interstellar medium
(ISM) to describe star formation and supernova feedback 
\citep{springel:multiphase,springel:models}.
Feedback from supernovae is captured in this sub-resolution model
through an effective equation of state for star-forming gas, enabling
us to stably evolve disks with arbitrary gas fractions \citep[see][]{springel:spiral.in.merger,
robertson:disk.formation}. This feedback prescription can be adjusted 
between an isothermal gas with effective
temperature of $10^4$ K and our full multiphase model
with an effective temperature $\sim10^5$ K.

Supermassive black holes are represented by ``sink'' particles
that accrete gas at a rate $\dot{M}$ estimated from the local gas
density and sound speed using an Eddington-limited prescription based
on Bondi-Hoyle-Lyttleton accretion theory.  The bolometric luminosity
of the black hole is taken to be $L_{\rm bol}=\epsilon_{r}\dot{M}\,c^{2}$,
where $\epsilon_r=0.1$ is the radiative efficiency.  We assume that a
small fraction (typically $\approx 5\%$) of $L_{\rm bol}$ couples dynamically
to the surrounding gas, and that this feedback is injected into the
gas as thermal energy, weighted by the SPH smoothing kernel.  This
fraction is a free parameter: we adjust it to match the normalization of 
the local $M_{\rm BH}-\sigma$ relation as in  \citet{dimatteo:msigma,
dimatteo:cosmo.bhs,sijacki:radio}. 
We emphasize that this controls only the normalization of this relation; 
i.e.\ inefficient feedback coupling means a BH must grow proportionally larger in order to couple 
the same energy to the ISM and self-regulate, but the scalings of 
BH mass with $\sigma$ and host properties (i.e.\ slopes of the 
BH-host relations and correlations between residuals in these 
relations) are {\em not} changed. Because our 
comparisons throughout are based on the relative scalings of BH mass 
with host properties, this normalization choice is simply a matter of convenience. 
For now, we do not resolve the small-scale dynamics of the gas in the immediate
vicinity of the black hole, but assume that the time-averaged
accretion rate can be estimated from the gas properties on the scale
of our spatial resolution (roughly $\approx 20$\,pc, in the best
cases).

The progenitor galaxies in the mergers are constructed following
\citet{springel:models}.  For each
simulation, we generate two stable, isolated disk galaxies, each with
an extended dark matter halo with a \citet{hernquist:profile} profile,
motivated by cosmological simulations \citep[e.g.,][]{nfw:profile,busha:halomass}, 
an exponential disk of gas and stars, and (optionally) a
bulge.  The galaxies have total masses $M_{\rm vir}=V_{\rm
vir}^{3}/(10GH_{0})$ for $z=0$, with the baryonic disk having a mass
fraction $m_{\rm d}=0.041$, the bulge (when present) having $m_{\rm
b}=0.0136$, and the rest of the mass in dark matter.  The dark matter
halos are assigned a
concentration parameter scaled as in \citet{robertson:msigma.evolution} appropriately for the 
galaxy mass and redshift following \citet{bullock:concentrations}. 
The disk scale-length is computed
based on an assumed spin parameter $\lambda=0.033$, chosen to be near
the mode in the $\lambda$ distribution measured in simulations \citep{vitvitska:spin},
and the scale-length of the bulge is set to $0.2$ times this.

Typically, each galaxy initially consists of 600,000 dark matter halo
particles, 20,000 bulge particles (when present), 40,000 gas and 40,000
stellar disk particles, and one BH particle.  We vary the numerical
resolution, with many simulations using twice, and a subset up to 128
times, as many particles. We choose the initial seed
mass of the black hole either in accord with the observed $M_{\rm
BH}$-$\sigma$ relation or to be sufficiently small that its presence
will not have an immediate dynamical effect, but we have varied the seed
mass to identify any systematic dependencies.  Given the particle
numbers employed, the dark matter, gas, and star particles are all of
roughly equal mass, and central cusps in the dark matter and bulge
are reasonably well resolved \citep[see Figure 2 in][]{springel:models}.

We consider the set of several hundred simulations from 
\citet{robertson:fp}, in which we vary the numerical resolution, the orbit of the
encounter (disk inclinations, pericenter separation), the masses and
structural properties of the merging galaxies, initial gas fractions,
halo concentrations, the parameters describing star formation and
feedback from supernovae and black hole growth, and initial black hole
masses. The detailed list of varied properties is given in Tables~1 \&\ 2 
of \citet{robertson:fp}. For example, 
the progenitor galaxies have virial velocities $\vvir=50, 80, 115, 160,
226, 320,$ and $500\,{\rm km\,s^{-1}}$, and are constructed to match disks 
at redshifts $z=0, 2, 3, {\rm
and}\ 6$, and our simulations span a range in final black hole mass
$\mbh\sim10^{5}-10^{10}\,M_{\sun}$.  The extensive range of conditions
probed provides a large dynamic range, with final spheroid masses
spanning $\mstar\sim10^{8}-10^{13}\,M_{\sun}$, covering the
entire range of the observations we consider at all redshifts, and
allows us to identify any systematic dependencies in our models.  We
consider initial disk gas fractions (by mass) of $\fgas = 0.05,\ 0.2,\ 0.4,\ 0.6,\ 
0.8,\ {\rm and}\ 1.0$ for several choices of virial velocities,
redshifts, and ISM equations of state.  

The results described in this
paper are based primarily on simulations of equal-mass mergers;
however, by examining a small set of simulations of unequal mass
mergers, we find that the behavior does not change significantly for
mass ratios down to about 1:3 or 1:4, below which mass ratio the mergers 
produce neither substantial BH nor bulge growth, and therefore 
the are no longer appropriate to compare to local relations between 
BHs and massive spheroids.

\subsection{Analysis}
\label{sec:sims:analysis}

Each simulation is evolved until the merger is complete and the remnants are 
fully relaxed, typically $\sim1-2$\,Gyr after the final merger 
and coalescence of the BHs. We then measure kinematic properties of the 
remnants following \citet{robertson:fp,cox:kinematics}. 
The effective radius $\re$ is the projected half-mass stellar 
effective radius, and the velocity dispersion $\sigma$ is the average 
one-dimensional velocity dispersion within a circular 
aperture of radius $\re$. Projected quantities such as $\re$, $\sigma$, and 
the stellar surface mass density $I_{e}\equiv M_{\ast}(r < \re)/\pi\re^{2}$ 
are averaged over 100 random lines of sight to the remnant. 
Throughout, the stellar mass $M_{\ast}$ refers to the total stellar mass of the galaxy, and 
the dynamical mass $\mdyn$ refers to the 
traditional dynamical mass estimator 
\begin{equation}
\mdyn\equiv k\,\frac{\sigma^{2}\,\re}{G},
\end{equation}
where we adopt $k=8/3$ (although this choice is irrelevant as long as we apply it 
uniformly to both observations and simulations). We define 
$\phi_{c}$ as the gravitational potential from the 
galaxy (excluding the BH itself) at the location of the BH 
\citep[given the typical resolution over which this quantity is 
smoothed, this is effectively the 
converged bulge potential at $r=0$, in the absence of the BH; see][]{springel:gadget}. 
As a concentration index we 
adopt the ratio of half-mass radius $\re=R_{50}$ to $30\%-$mass radius $R_{30}$, 
and measure $n_{s}$ from projected mock images following Krause et al. (2007,
in preparation). 
We note that, for convenience, $\fgas$ typically refers to the gas fraction in the merging 
disks when the simulations are initialized, but show in \S~\ref{sec:driving} that 
our results are unchanged (although $\fgas$ itself systematically shifts) regardless of 
the time before the merger at which we choose to define the gas fraction of the 
systems.

\section{The Local BH-Host Correlations}
\label{sec:local}

\subsection{One-to-One Relationships}
\label{sec:local:onetoone}

\begin{figure}
    \centering
    \figexpand
    \plotone{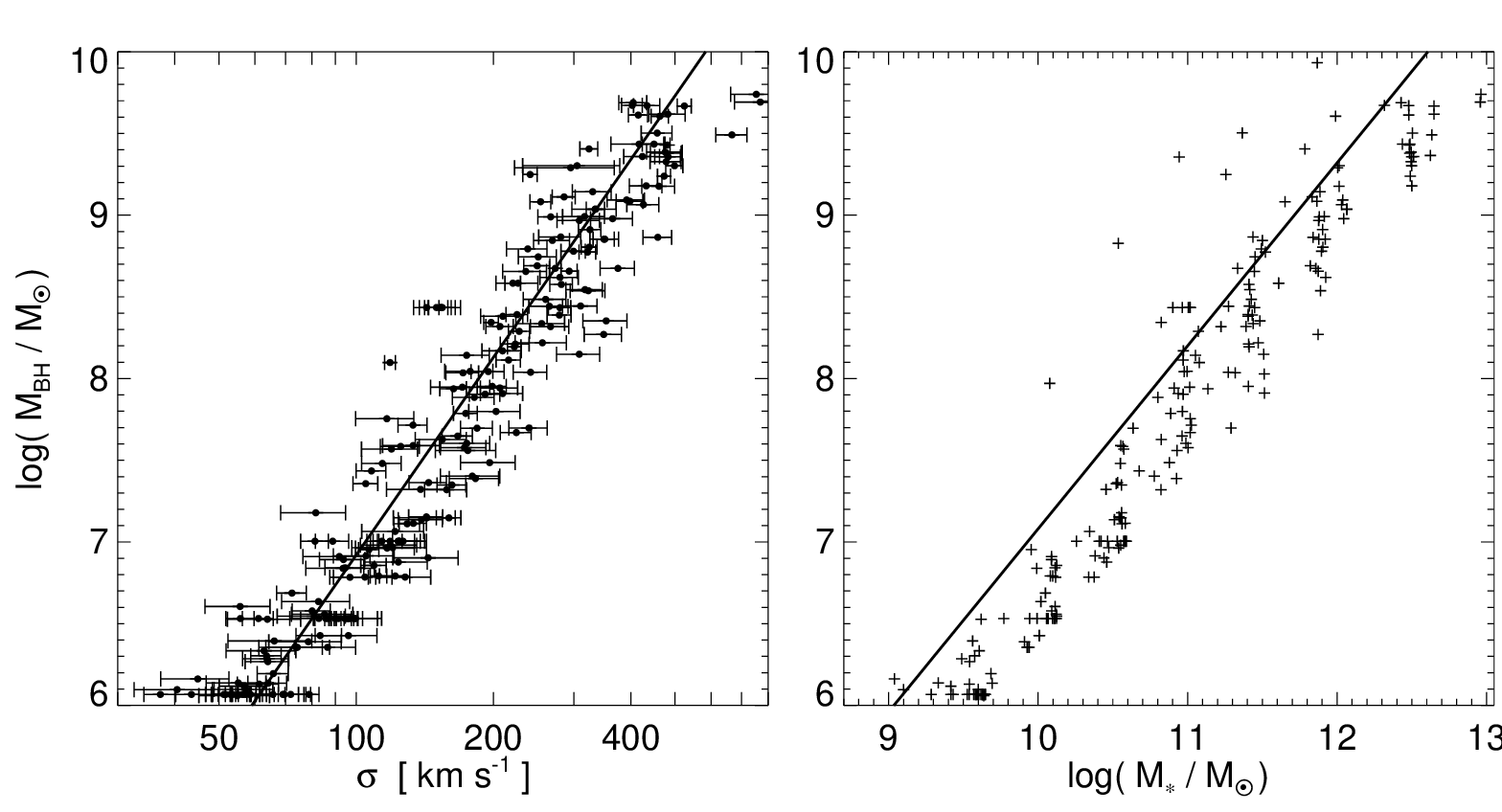}
    \caption{Location of our simulation merger remnant spheroids on the 
    $\mbh-\sigma$ and $\mbh-\mstar$ relations 
    \citep[as in][]{dimatteo:msigma,robertson:msigma.evolution}. 
    Solid lines show the 
    observed relations from \citet{tremaine:msigma} and \citet{haringrix}. 
    Error bars in $\sigma$ show the dispersion across 100 random viewing 
    angles. The simulations reproduce well the observed relations over a 
    wide dynamic range. As discussed in the text, there 
    are a number of likely reasons for the slight ($0.2\,$dex) 
    normalization offset in $\mstar$, but insofar as the slopes of the relations are 
    identical, this has no effect on our analysis. 
    \label{fig:msigma.demo}}
\end{figure}

Figure~\ref{fig:msigma.demo} shows the location of our simulation remnants on 
the the $\mbh-\sigma$ and $\mbh-\mstar$ relations. As demonstrated by 
\citet{dimatteo:msigma}, they agree well with the observed relations over 
a large dynamic range. Critically, although modifying our feedback prescriptions 
and, as we show below, adjusting the kinematic properties of the remnants 
by changing e.g.\ orbital parameters and gas fractions of the merging systems 
can shift the normalization of the relations, the slopes are {\em not} adjustable or 
tunable, but are a natural consequence of self-regulated BH growth. We note that 
there is a slight offset between the normalization (but not slope) of 
our predicted $\mbh-\mstar$ relation and that measured 
by \citet{haringrix}, but a weaker offset in $\mbh-\sigma$. 
This owes to the fact that, at fixed $\sigma$, our simulated systems typically have slightly 
larger (mean offset $\approx 0.16\,$dex) stellar masses than the observed 
systems on the $\mbh-\sigma$ relation. There are a number of 
possible explanations for this offset in the Faber-Jackson ($M_{\ast}[\sigma]$) relation. 
As pointed out by \citet{bernardi:magorrian.bias}, 
the systems with measured BH masses in \citet{haringrix} 
actually lie above the Faber-Jackson relation observed for typical early-type galaxies, perhaps 
owing to a selection bias. Their estimate of the magnitude of this bias is quite similar 
to the offset here and can completely account for our results.

Furthermore, we show below 
that at fixed $\mstar$, changing the gas fractions of the merging systems, their orbits, or 
their structural properties can systematically drive changes in $\sigma$. This means that  
the precise normalization of the observed Faber-Jackson relation depends, in detail, on the exact star 
formation and merger histories of the systems observed. Since we do not model these cosmological 
histories, but rather isolate different mergers in order to study how these changes are driven, 
it is not surprising that the normalizations of the relations do not happen to match perfectly. In any case, 
we are interested in how offsets or evolution from one relation or the other are produced, and 
how such residuals may scale with other host properties, for which the actual normalization of 
the relation factors out completely. It therefore makes no difference to our analysis (although 
we have considered both cases) whether we compare to our full suite or select only a subset of  
simulations which reproduce the (mean) normalization in the observed Faber-Jackson relation.

\begin{figure*}
    \centering
    \plotone{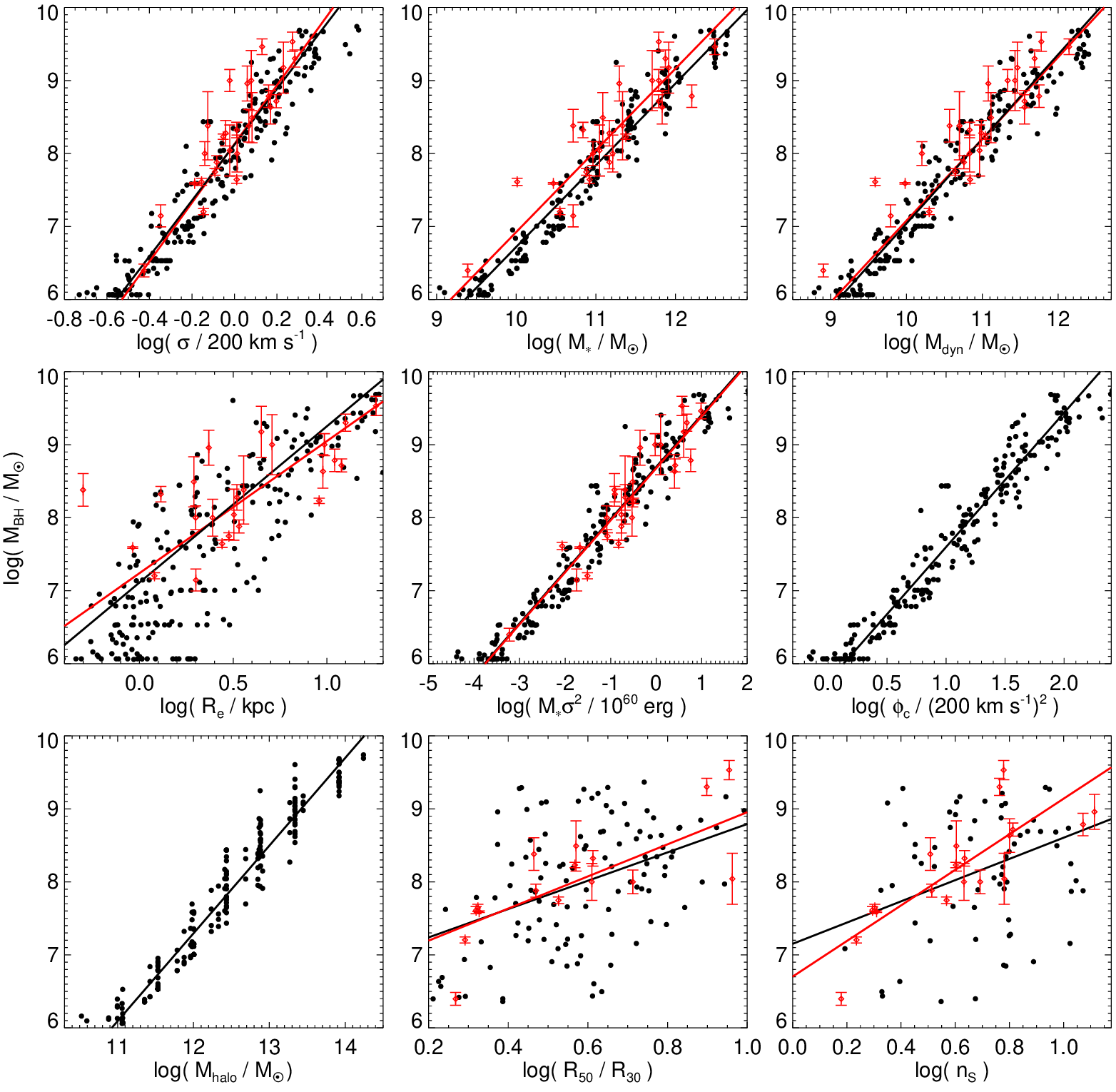}
    \caption{Correlations between BH mass and a variety of host spheroid properties. 
    Red points with error bars are the observations; red line shows the 
    least-squares best-fit to each relation. 
    Note that the quality of the correlation between $\mbh$ and concentration or 
    Sersic index observed is significantly reduced relative to that 
    in \citet{graham:concentration,graham:sersic} when we adopt the 
    Sersic index measurements from \citet{kormendy:wetvsdry}.
    Black points and lines are the 
    results of hydrodynamic simulations (see \S~\ref{sec:sims}). The slopes of the 
    simulated and observed relations 
    are statistically identical in every case, with all normalization offsets owing to the 
    small Faber-Jackson ($\mstar(\sigma)$) offset in Figure~\ref{fig:msigma.demo}. 
    \label{fig:all.correlations}}
\end{figure*}

Figure~\ref{fig:all.correlations} shows the correlation between BH mass 
and a wide variety of host properties, from both our simulations and the observed 
sample. The slopes of the simulated correlations are essentially identical to those 
observed in every case. Note that many of the correlations are similarly tight, 
including the correlations with velocity dispersion $\sigma$, stellar mass 
$\mstar$, dynamical mass $\mdyn$, effective bulge binding energy $\mstar\,\sigma^{2}$, 
and central potential $\phi_{c}$. The best-fit correlations are listed in 
Table~\ref{tbl:correlations}, along with the intrinsic scatter in $\mbh$ 
estimated about each correlation from the simulations.
We do not list the correlation with $M_{\rm halo}$, as 
it is clear in our simulations that BH mass is correlated with small-scale bulge 
properties (unsurprising, given that the central potential of the bulge is 
strongly baryon-dominated). Therefore, while there is an indirect correlation with $M_{\rm halo}$ 
through e.g.\ the $M_{\rm halo}-\mstar$ and $\mbh-\mstar$ relations, its nature 
depends systematically on the exact $M_{\rm halo}-\mstar$ relation.

\tableclear
%\begin{landscape}
%\begin{deluxetable*}{cccccccc}
\begin{\tableset}{cccccccc}
%\rotate
%\tablecolumns{13}
\tabletypesize{\scriptsize}
\tablecaption{BH-Host Correlations\label{tbl:correlations}}
\tablewidth{0pt}
\tablehead{
\multicolumn{1}{c}{} & 
\multicolumn{3}{c}{Observed} & 
\multicolumn{1}{c}{} & 
\multicolumn{3}{c}{Simulated} \\
\cline{2-4}\cline{6-8}\\
\colhead{Variables\tablenotemark{1,5}} &
\colhead{Normalization\tablenotemark{2}} &
\colhead{Slope\tablenotemark{3}} &
\colhead{Scatter\tablenotemark{4}} &
\colhead{} &
\colhead{Normalization} &
\colhead{Slope\tablenotemark{3}} &
\colhead{Scatter}
}
\startdata
%$\sigma^{\alpha}\,\re^{\beta}$ & $8.04$ & $3.02,\ 0.55$ & $0.19$ & 
%$8.36\pm0.06$ & $2.72\pm0.24,\ 0.54\pm0.13$ & $0.21$ \\ (just MH03 values for Re)
$\sigma^{\alpha}\,\re^{\beta}$ & $8.33\pm0.06$ & $3.00\pm0.30,\ 0.43\pm0.19$ & $0.21$ &  $ $ & 
$8.16\pm0.05$ & $2.90\pm0.38,\ 0.54\pm0.11$ & $0.21$ \\ %(updated values for Re, sigma)
\\
%$\mstar^{\alpha}\,\sigma^{\beta}$ & $7.90\pm0.06$ & $0.74\pm0.17,\ 1.86\pm0.51$ & $0.19$ & $ $ & 
$\mstar^{\alpha}\,\sigma^{\beta}$ & $8.24\pm0.06$ & $0.54\pm0.17,\ 2.18\pm0.58$ & $0.22$ & $ $ & 
$7.93\pm0.06$ & $0.72\pm0.12,\ 1.40\pm0.49$ & $0.19$ \\
\\
$\mstar^{\alpha}\,\re^{\beta}$ & $8.06\pm0.07$ & $1.78\pm0.40,\ -1.05\pm0.37$ & $0.25$ &  $ $ & 
$7.64\pm0.04$ & $1.50\pm0.22,\ -0.56\pm0.26$ & $0.21$ \\ %(updated values for Re, sigma)
\\
$\mstar\,\sigma^{2}$ & $8.23\pm0.06$ & $0.71\pm0.06$ & $0.25$ &  $ $ & 
$7.92\pm0.04$ & $0.71\pm0.03$ & $0.21$ \\
\\
$\sigma$ & $8.28\pm0.08$ & $3.96\pm0.39$ & $0.31$ &  $ $ & 
$8.04\pm0.06$ & $3.91\pm0.20$ & $0.31$ \\
\\
$\mstar$ & $8.21\pm0.07$ & $0.98\pm0.10$ & $0.33$ &  $ $ & 
$7.85\pm0.05$ & $1.16\pm0.06$ & $0.23$ \\
\\
$\mdyn$ & $8.22\pm0.10$ & $1.05\pm0.13$ & $0.43$ &  $ $ & 
$8.18\pm0.06$ & $1.05\pm0.06$ & $0.28$ \\
\\
$\re$ & $8.44\pm0.10$ & $1.33\pm0.25$ & $0.45$ &  $ $ & 
$8.48\pm0.12$ & $1.92\pm0.14$ & $0.56$ \\
\\
%$M_{\rm halo}$ & $8.52\pm0.05$ & $1.16\pm0.06$ & $0.24$ &  $ $ & -- & -- & -- \\
$\phi_{\rm c}$ & -- & -- & --  &  $ $ & $8.31\pm0.05$ & $1.77\pm0.08$ & $0.23$\\
\enddata
\tablenotetext{1}{For the variables $(x,\ y)$, a 
correlation of the form $\log(\mbh)=\alpha\log(x)+\beta\log(y)+\delta$ is assumed, 
where the normalization is $\delta$ and $\alpha$, $\beta$ are the logarithmic slopes.}
\tablenotetext{2}{The normalization gives $\log(\mbh/M_{\sun})$ for $\sigma=200\,{\rm km\,s^{-1}}$, 
$\mstar=10^{11}\,M_{\sun}$, $\mdyn=10^{11}\,M_{\sun}$, $\re=5\,{\rm kpc}$, 
which roughly minimizes the covariance between fit parameters.}
\tablenotetext{3}{Errors quoted here for the BHFP relations in $(\sigma,\ \re)$, $(\mstar,\ \sigma)$, and 
$(\mstar, \re)$ include the covariance between the two slopes. Holding one of the two fixed and 
varying the other yields substantially smaller errors (typically $\sim5\%$). All quoted errors 
account for measurement errors in both $\mbh$ and the relevant independent variables.}
\tablenotetext{4}{The internal scatter is estimated from both the simulations and observations 
as that which yields a reduced $\chi^{2}/\nu=1$ with respect to the given best-fit relation.}
\tablenotetext{5}{Central potential $\phi_{\rm c}$, normalization at 
$\phi_{\rm c}=10^{6}\,{\rm km^{2}\,s^{-2}}$. There are no observational 
measurements presently available to compare with this correlation.}
%\end{deluxetable*}
\end{\tableset}
%\end{landscape}
\tableclear

We also note that the (relatively large) 
scatter in the correlations between $\mbh$ and concentration or 
Sersic index in Figure~\ref{fig:all.correlations} 
appears contrary to the conclusions of \citet{graham:concentration}, 
who argue for very small intrinsic scatter in these correlations.
However, \citet{novak:scatter} point out that uncertainty in this correlation, 
unlike for the $\mbh-\sigma$ or 
$\mbh-\mdyn$ relations, is dominated by the measurement errors in concentration index 
or $\sersic$, which means that improved observations are needed to determine whether 
the relation is actually consistent with small intrinsic scatter. In fact, when we update the measurements 
from \citet{graham:concentration} and \citet{graham:sersic} with the $\sersic$ measurements from 
\citet{kormendy:wetvsdry}, which typically reduce the measurement 
error in $\sersic$ from 
$\sim20\%$ to $<5\%$ (and in at least two cases\footnote{
\citet{graham:sersic} quote $\sersic=3.04^{+0.61}_{-0.51},\ 2.73^{+0.55}_{-0.46},\ $ 
for NGC 3377 and 4473, respectively, whereas 
\citet{kormendy:wetvsdry} measure $\sersic=4.917,\ 6.04\pm0.33$. The 
newer values make the measured $\mbh$ discrepant with the $\mbh-\sersic$ relation 
in \citet{graham:sersic} by $0.64$ and $0.78$\,dex, respectively (with $\approx0.15$\,dex 
measurement errors in $\mbh$ for each).} 
change $\sersic$ by $>3\sigma$ 
relative to the \citet{graham:sersic} fit), the quality of the correlation is substantially degraded, and 
a significantly larger intrinsic scatter is implied. 

In the simulations, it is possible, given appropriate 
gas fractions or orbital parameters, to substantially change $\sersic$ or $R_{30}/R_{50}$ 
at fixed stellar mass without driving a corresponding change in 
$\mbh$ ($\sersic$ also appears to be more variable sightline-to-sightline than 
other projected quantities such as $\sigma$ or $\re$). 
As will be discussed in detail in Krause et al.\ (2007, in preparation), 
this has the consequence that the values of $\sersic$ predicted by 
a uniform set of simulations are not as tightly correlated with spheroid 
mass as is observed (and, as a secondary consequence, $\mbh$, which 
is tightly correlated with $\mstar$, is less well-correlated with $\sersic$). 
Furthermore, the weaker correlation in the simulations 
should not be surprising -- unlike the $\mstar-\re$ or 
$\mstar-\sigma$ correlations which are produced fundamentally (at least to 
lowest order) by basic dynamical effects (and are not especially sensitive to the 
exact type of spheroid-producing mergers), it is commonly believed 
\citep[e.g.,][]{kormendy:wetvsdry} that the correlation between $\sersic$ and $\mstar$ 
is driven by an increasing prevalence of dissipationless (spheroid-spheroid) 
mergers at higher masses. Such mergers will, by definition, conserve the $\mbh-\mstar$ relation, 
but simulations have shown that they dramatically change $\sersic$ 
\citep[e.g.][]{boylankolchin:dry.mergers} and therefore the $\mbh-\sersic$ relation. 
If this is true, then the 
$\mbh-\sersic$ correlation, unlike the $\mbh-\mstar$ relation, must fundamentally be driven 
by cosmological effects (e.g.\ the differential contribution of different merger types 
as a function of mass and redshift), which our simulations do not represent. A 
more accurate modeling is outside the scope of this paper, as it 
would require a fully cosmological prediction for 
merger rates (as a function of type, mass, and redshift) 
with high-resolution cosmological 
simulations such that the central galaxy structure and 
BH feedback effects could be modeled. The $\mbh-\mstar$ and BHFP relations 
on which we focus, on the other hand, should be relatively robust to these 
effects (see also \S~\ref{sec:discussion}).

\subsection{A Black Hole Fundamental Plane}
\label{sec:local:FP}

\begin{figure}
    \centering
    \figexpand
    \plotter{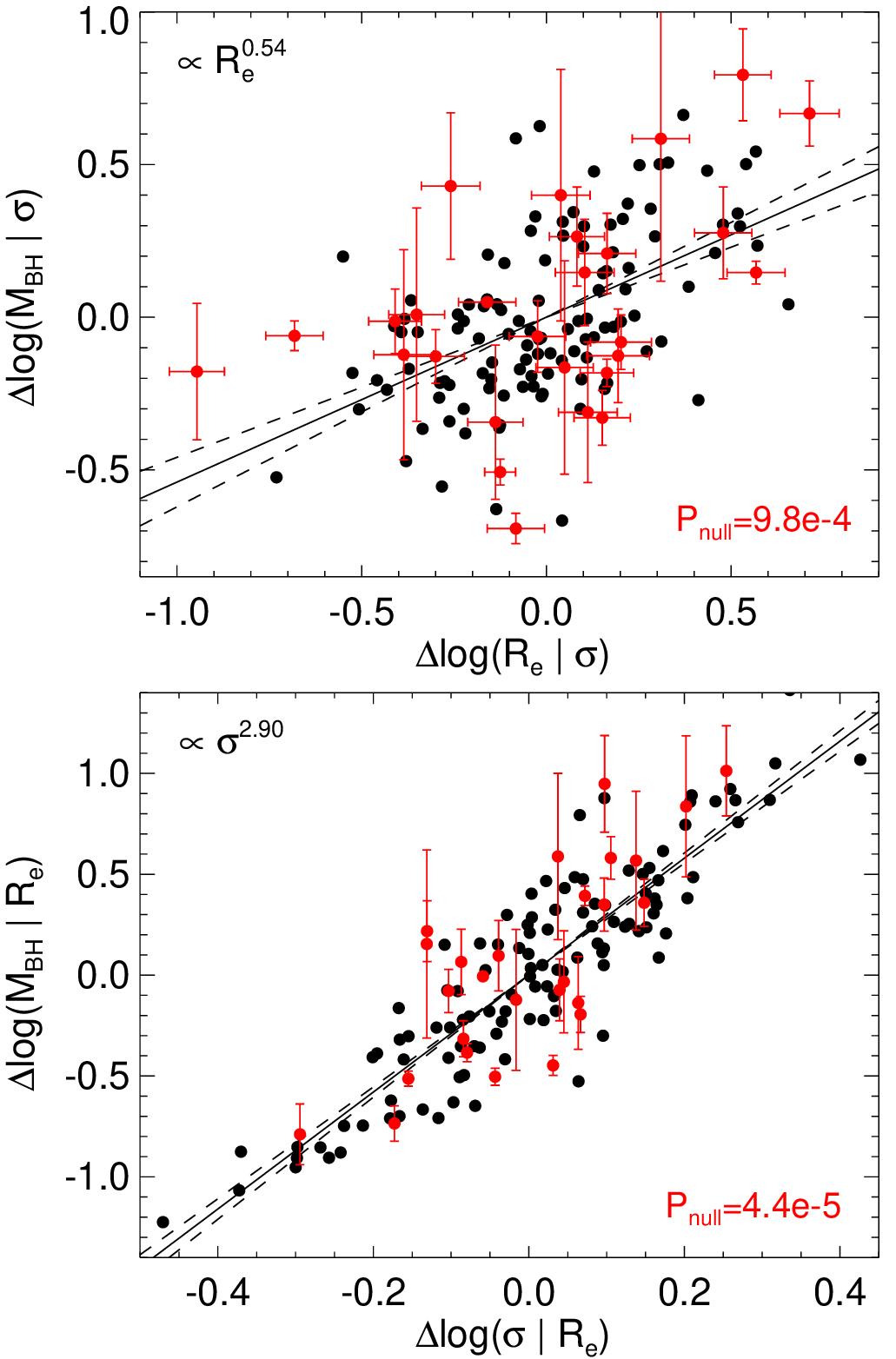}
    \caption{{\em Upper:} Correlation between the residuals in the $\mbh-\sigma$ 
    relation and $\re-\sigma$ relation, from our simulations (black points) and 
    observed sample (red points with errors). At fixed $\sigma$, systems with larger 
    effective radii $\re$ also have larger black hole masses $\mbh$. The fit to this residual correlation is 
    shown with the black lines ($\pm1\,\sigma$ range in the best-fit correlation shown 
    as dashed lines -- note that they are strongly inconsistent with zero correlation), 
    with the slope shown. The probability of the null hypothesis of no correlation in the 
    residuals (i.e.\ no systematic dependence of $\mbh$ on $\re$ at fixed $\sigma$) for 
    the observed systems is shown (red $P_{\rm null}$) -- the observations imply 
    a secondary ``fundamental plane''-type correlation at $3\,\sigma$. {\em Lower:} Same, 
    but considering the correlation between $\mbh$ and $\sigma$ at fixed effective radius 
    $\re$. 
    \label{fig:res.reff.sigma}}
\end{figure}

We wish to determine whether or not a simple one-to-one correlation between e.g.\ 
$\mbh$ and $\sigma$ is a sufficient description of the simulations, or 
whether there is evidence for additional dependence on a second parameter such as 
$\re$ or $\mstar$. The most efficient 
way to determine such a dependence is by looking for correlations 
between the 
residuals of the various projections of such a potential BHFP relation. 
Following \paperone, Figure~\ref{fig:res.reff.sigma} plots the correlation between 
BH mass $\mbh$ and host 
bulge effective radius $\re$ at fixed $\sigma$. Specifically, we determine the 
residual with respect to the $\mbh-\sigma$ relation by 
fitting $\mbh(\sigma)$ to an arbitrary log-polynomial 
\begin{equation}
\langle\log(\mbh)\rangle=\Sigma\,{\bigl[}\,a_{n}\,\log(\sigma)^{n}{\bigr]}, 
\end{equation}
allowing as many terms as the data favor (i.e.\ until $\Delta\chi^{2}$ with respect to 
the fitted relation is $<1$), and then taking 
\begin{equation}
\Delta\log(\mbh\,|\,\sigma)\equiv \log(\mbh) - \langle\log(\mbh)\rangle(\sigma).
\end{equation}
We determine the residual $\Delta\log(\re\,|\,\sigma)$
(or $\Delta\log(\re\,|\,\sigma)$, for the stellar mass) in identical fashion, and 
plot the correlation between the two. 
We allow arbitrarily high terms in $\log(\sigma)$ to avoid introducing bias 
by assuming e.g.\ a simple power-law correlation between $\mbh$ and $\sigma$, but 
find in practice that such terms are not needed -- as discussed in \paperone, there is 
no significant evidence for a log-quadratic (or higher-order) dependence 
of $\mbh$ on $\sigma$, $\re$, or $\mstar$, so allowing for these terms 
changes the residual best-fit solutions by $\ll 1\sigma$. 
Of course, even this approach could 
in principle introduce a bias via our assumption of some functional form, and 
so we have also considered a non-parametric approach where 
we take the mean $\langle \log(\mbh) \rangle$ in bins of $\log(\sigma)$. 
Our large set of simulations allows us to do this with very narrow binning,
and we recover a nearly identical answer.  

\begin{figure}
    \centering
    \figexpand
    \plotter{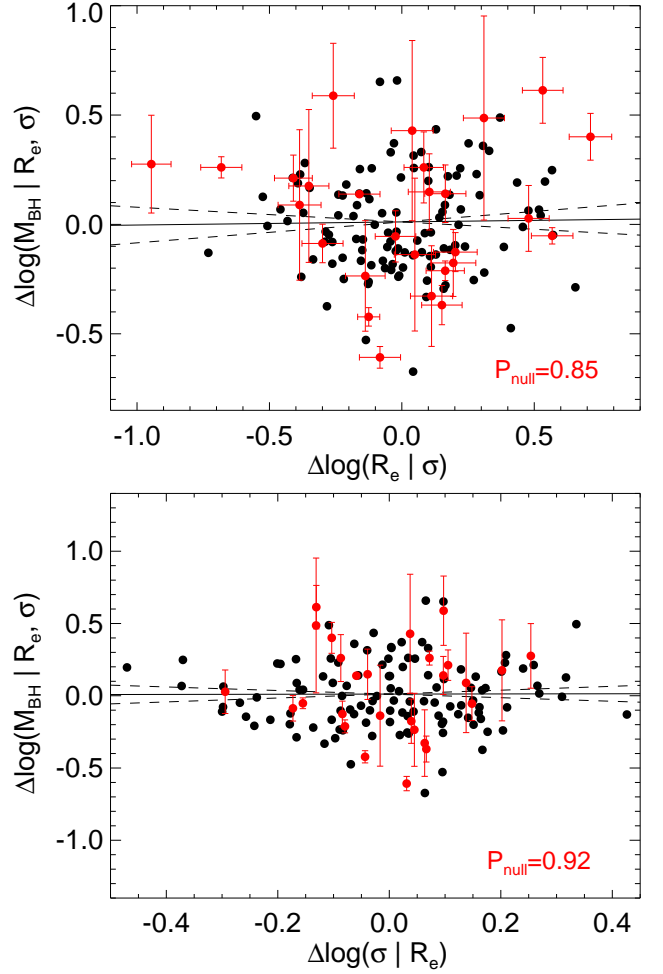}
    \caption{Correlation between the residuals in our BH ``fundamental plane'' relation 
    $\mbh\propto\sigma^{2.90}\,\re^{0.54}$ 
    and effective radius $\re$ at fixed $\sigma$ ({\em upper}) or $\sigma$ at fixed 
    $\re$ ({\em lower}). Accounting for the joint dependence of $\mbh$ on 
    $\sigma$ and $\re$ removes the strong systematic dependencies in the residuals 
    from Figure~\ref{fig:res.reff.sigma} ($P_{\rm null}$ is large, meaning there is no 
    further residual dependence). 
    \label{fig:res.reff.sigma.fp}}
\end{figure}

The simulations show a highly significant correlation between 
$\mbh$ and $\re$ at fixed $\sigma$, similar to the observed trend in residuals. 
We therefore 
introduce a FP-like relation of the form 
\begin{equation}
\mbh\propto\sigma^{\alpha}\,\re^{\beta} \, ,
\end{equation}
which can account for these dependencies. 
Formally, we determine the combination of $(\alpha,\,\beta)$ which simultaneously 
minimizes the $\chi^{2}/\nu$ of the fit and 
the significance of the correlations between the residuals in $\sigma$ and 
$\mbh$ (or $\re$ and $\mbh$). This yields similar results to 
the direct fitting method of 
\citet{bernardi:fp} from the spheroid FP, which minimizes 
\begin{equation}
\Delta^{2} = {\bigl[} \log(\mbh) - \alpha\log(\sigma) - \beta\log(\re) - \delta {\bigr]}^{2}, 
\end{equation}
and corresponds exactly to the method used in fitting the observations in 
\paperone, to which we compare.
This yields a best-fit BHFP relation 
\begin{eqnarray}
\label{eqn:fp.sigma.reff}
  \log(\mbh) &=& 8.16 + 2.90(\pm0.38)\,\log(\sigma/200\,{\rm km\,s^{-1}}) \\
\nonumber & & + 0.54(\pm0.11)\,\log(\re/5\,{\rm kpc}) 
\end{eqnarray}
from the simulations, similar to the observed BHFP relation, 
$\mbh\propto \sigma^{3.0\pm0.3}\,\re^{0.43\pm0.19}$ (\paperone). 
Unsurprisingly, the slopes in the BHFP relation are close to those formally determined for the 
residuals in Figure~\ref{fig:res.reff.sigma}. 
Figure~\ref{fig:res.reff.sigma.fp} plots the 
residuals of $\mbh$ with respect to these 
fundamental plane relations, at fixed $\re$ and fixed $\sigma$. The introduction of a 
BHFP eliminates the strong systematic correlations between the residuals, yielding 
flat errors as a function of $\sigma$ and $\re$. 

Given the definition of $\mdyn\propto\sigma^{2}\,\re$, it is trivial to convert 
the best-fit  
BHFP relation in terms of $\sigma$ and $\re$ 
to one in $\mdyn$, obtaining
\begin{eqnarray}
  \log(\mbh) &=& 8.11 + 0.54\,\log(\mdyn/10^{11}\,M_{\sun}) \\
\nonumber & & + 1.82\,\log(\sigma/200\,{\rm km\,s^{-1}}) \\ 
\nonumber &=& 8.03 + 1.45\,\log(\mdyn/10^{11}\,M_{\sun}) \\
\nonumber & & - 0.91\,\log(\re/5\,{\rm kpc})  \, .
\end{eqnarray}
As is the case for the observed systems, the correlation with either $\sigma$ or 
$\re$ at fixed $\mdyn$ is non-zero and highly significant (i.e.\ the BHFP 
is not simply a reflection of a simpler $\mbh-\mdyn$ relation).

\begin{figure}
    \centering
    \figexpand
    \plotter{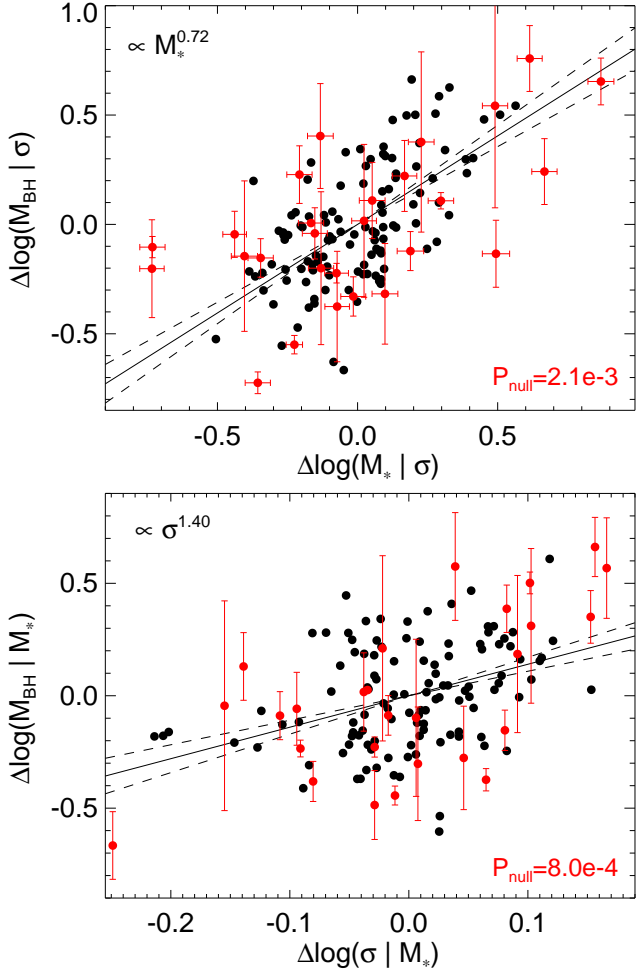}
    \caption{As Figure~\ref{fig:res.reff.sigma}, but considering the correlation between 
    the residual in BH mass $\mbh$ and stellar mass $\mstar$ at fixed velocity 
    dispersion $\sigma$ ({\em upper}), or between $\mbh$ and $\sigma$ at 
    fixed $\mstar$ ({\em lower}). Observed and simulated systems with larger velocity 
    dispersions (deeper potential wells) at fixed stellar mass have more massive BHs, as do  
    more massive systems at fixed $\sigma$. A nearly identical result is obtained 
    using dynamical mass $\mdyn$ instead of stellar mass $\mstar$: 
    a residual dependence $\propto\mdyn^{0.55\pm0.09}$ ($P_{\rm null}=0.0037$) at fixed 
    $\sigma$ and $\propto\sigma^{2.16\pm0.79}$ ($P_{\rm null}=0.0032$) at fixed 
    $\mdyn$.
    \label{fig:res.mstar.sigma}}
\end{figure}
\begin{figure}
    \centering
    \figexpand
    \plotter{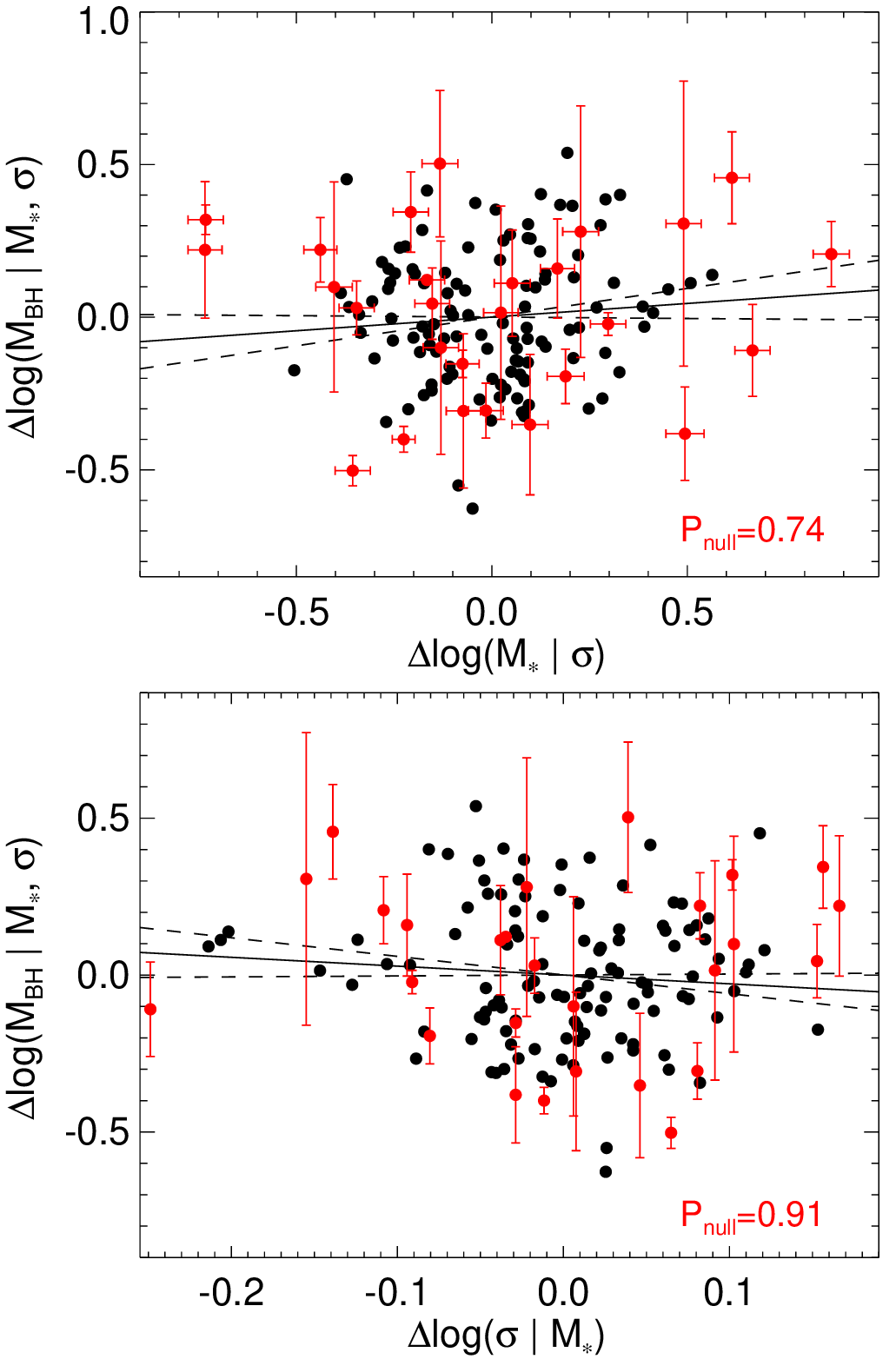}
    \caption{As Figure~\ref{fig:res.reff.sigma.fp}, but considering the residuals with respect the 
    the BHFP defined in terms of stellar mass and $\sigma$. Placing the simulations and 
    observations in the context of a FP relation eliminates the strong trends in 
    their residuals. 
    \label{fig:res.mstar.sigma.fp}}
\end{figure}

At low redshift, $\sigma$, $\re$, and $\mdyn$ can be determined reliably, but 
at high redshift it is typically the stellar mass $\mstar$ or luminosity which is used 
to estimate $\mbh$. Therefore, it is interesting to examine the BHFP projections 
in terms of e.g.\ $\mstar$ and $\sigma$ or 
$\mstar$ and $\re$. Repeating our analysis, we 
find in Figures~\ref{fig:res.mstar.sigma} \&\ \ref{fig:res.mstar.sigma.fp} 
that the observations demand a FP relation over a simple $\mbh(\mstar)$ relation at 
high significance. We find a nearly identical result using the dynamical mass, 
$\mdyn\propto\sigma^{2}\,\re$, instead of stellar mass $\mstar$, as 
expected from the BHFP in terms of $\sigma$ and $\re$. 
The exact values of the best-fit coefficients of this 
BHFP determined from the observations 
are given (along with those of various 
other BHFP projections) in Table~\ref{tbl:correlations}. 

Because the simulation merger remnant spheroids lie on a stellar-mass fundamental plane 
very similar to observed elliptical galaxies \citep{robertson:fp}, which tightly 
relates $\mstar$, $\re$, and $\sigma$, the BHFP in terms of any two of those variables can be 
easily converted to any other two. 
In other words, the 
two forms of the BHFP ($\mbh\propto \sigma^{3}\,\re^{1/2}$ and 
$\mbh\propto\sigma^{2}\,\mstar^{1/2}$) are completely equivalent (the choice 
between them is purely a matter of convenience), and it is redundant to 
search for a four-variable correlation 
(of the form $\mbh\propto\sigma^{\alpha}\,\re^{\beta}\,\mstar^{\gamma}$). As in 
the case of the observations (\paperone), we stress that 
no transformation (given the early-type FP relating these three 
variables) eliminates the dependence on two variables -- i.e.\ 
no transformation from any one BHFP allows us to write 
$\mbh$ as a pure function of either $\sigma$, $\re$, $\mstar$, 
or $\mdyn$.

\section{Implications of the BHFP for Local BH Masses and Demographics}
\label{sec:pred.local}

\begin{figure*}
    \centering
    \plotone{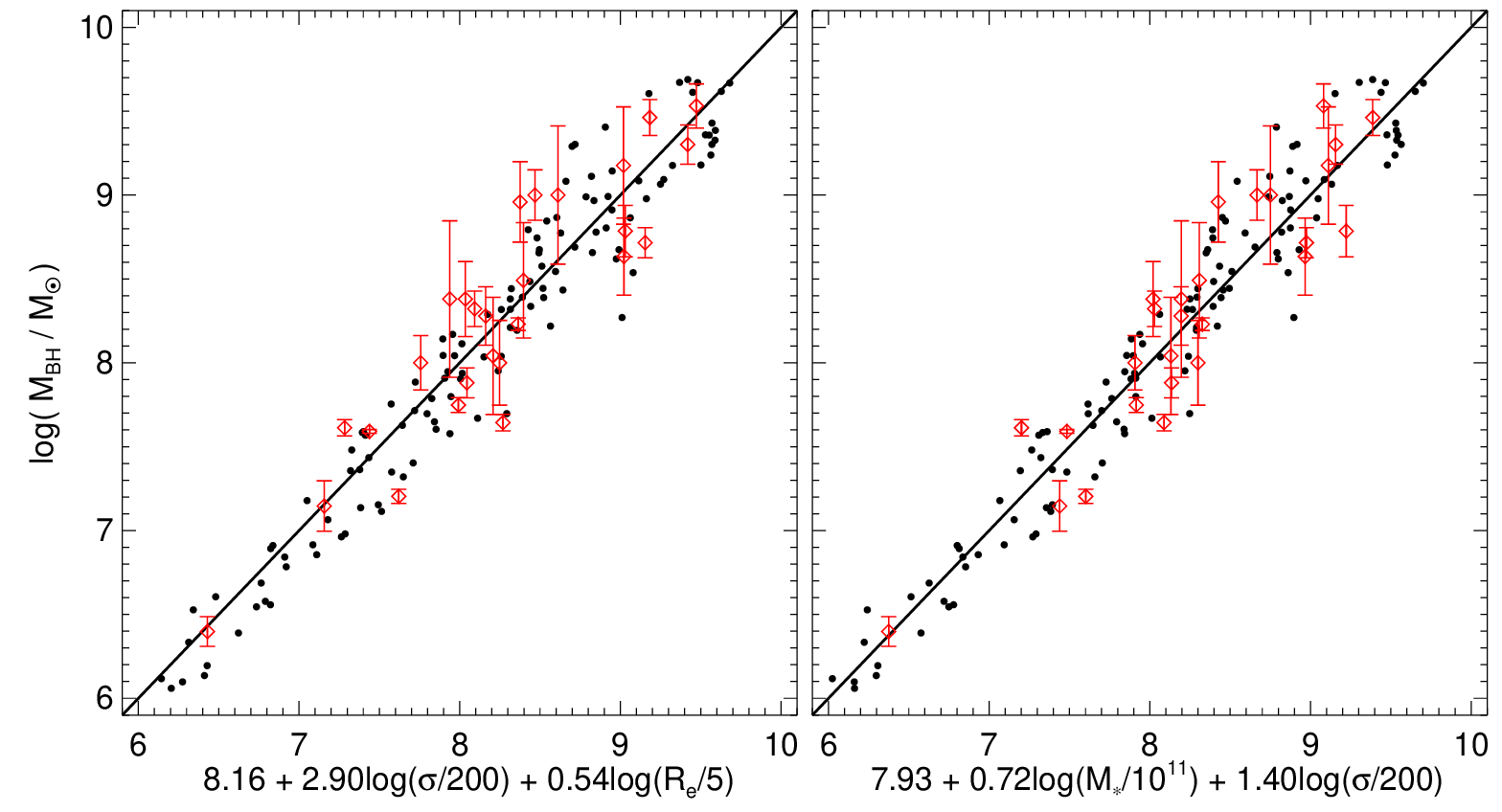}
    \caption{Masses of BHs in our simulations and from local measurements, 
    compared to the expectation from the best-fit BHFP relations in 
    $\sigma,\ \re$ and $\mstar,\ \sigma$. The two agree well at all masses, 
    without any evidence for curvature in the relations. The intrinsic scatter 
    in $\mbh$ at fixed $\sigma,\ \re$ or $\mstar, \sigma$ is 
    estimated from the simulations to be $\approx0.20\,$dex (see Table~\ref{tbl:correlations}), 
    which is consistent with the scatter in the observed points (given their measurement 
    errors). 
    \label{fig:show.local.fp}}
\end{figure*}

Given that the ``true'' correlation between $\mbh$ and host properties 
appears to follow a FP-like relation, it is natural to ask how adopting such a 
relation affects the estimation of BH masses from observed host properties. 
Figure~\ref{fig:show.local.fp} shows the observed and simulated systems in 
the fundamental plane. The relation appears to be a good predictor of 
$\mbh$ over a large dynamic range, and there is no evidence for any curvature 
or higher-order terms in the relation (fitting e.g.\ a log-quadratic relation 
in this space yields $\Delta \chi^{2}<1$). As detailed in Table~\ref{tbl:correlations}, 
the intrinsic scatter in the BHFP is small, typically $\sim0.2$\,dex, and in all cases 
smaller than the scatter in e.g.\ the $\mbh-\sigma$ or $\mbh-\mstar$ relations. 

However, 
as \citet{novak:scatter} note, minimizing the intrinsic scatter does not necessarily 
maximize the observational ability to predict BH masses. 
The BHFP relations depend on measuring two of either $\sigma$, $\re$, or $\mstar$, and 
therefore introduce additional errors from the measurements of two (as opposed to 
just one) of these quantities. At low 
redshifts, it may be possible to obtain accurate measurements of both $\sigma$ and 
$\re$ and therefore still obtain more accurate mass estimates (although we caution 
that several of the literature sources from which we compile observations differ by $>2\,\sigma$ 
in some $\re$ measurements, owing to various systematic issues such as the choice of observed 
bands). However, at high redshifts $\mstar$ remains the most easily applicable proxy 
for $\mbh$, and it is not clear that the additional accuracy gained by introducing 
the $\re$ term substantially improves the predictive power of the relation. 

Ultimately, the $\mbh-\sigma$ and $\mbh-\mstar$ relations are not much worse 
in a mean sense around 
$\mbh\sim10^{8}-10^{9}\,\msun$, 
with relatively small intrinsic scatter (see Table~\ref{tbl:correlations}). The 
reason these relations work as well as they do is that they are both 
nearly edge-on projections of the BHFP. Given a relation $\mbh\propto \sigma^{3}\,\re^{0.5}$, it 
is not surprising that $\sigma$ is an acceptable proxy for $\mbh$ in many situations (whereas 
the $\mbh-\re$ correlation has quite large scatter, as $\re$ enters with a relatively weak 
dependence in the BHFP). 

\begin{figure}
    \centering
    \figexpand
    \plotone{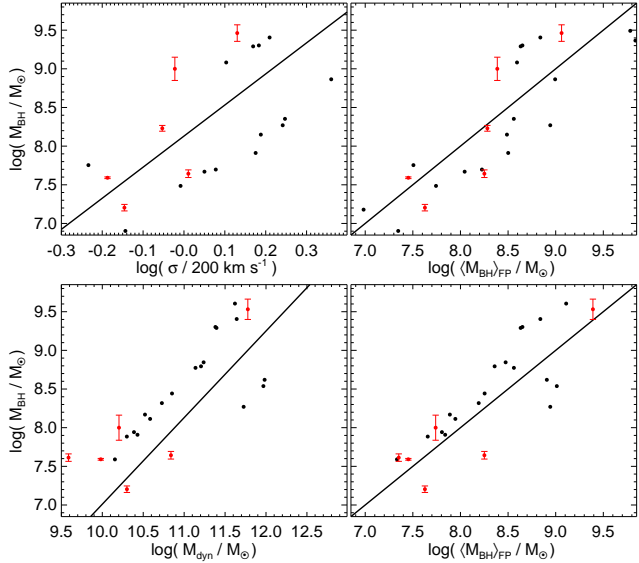}
    \caption{{\em Upper:} Outliers in the local $\mbh-\sigma$ relation 
    from the observations (red; defined as points $>3\,\sigma$ discrepant from 
    the mean trend) and simulations (black; defined as points $0.5\,$dex discrepant 
    from the mean trend) are plotted relative to the mean 
    relation from \citet{tremaine:msigma} (line) ({\em left}), and plotted relative 
    to the expected BH mass from the fundamental plane relation ({\em right}). 
    {\em Lower:} Same, but for outliers from the local $\mbh-\mdyn$ relation 
    from \citet{haringrix}. Most outliers in $\mbh-\sigma$ or $\mbh-\mdyn$ are 
    explained by the BHFP relations we derive -- i.e.\ they have abnormal 
    values of $\sigma$ or $\re$ for their masses, but are not outliers in the BHFP relation. 
    There are a small number of observed and simulated systems which are genuine outliers 
    from all relations -- i.e.\ do have anomalous BH masses, but we note that there are 
    no simulated or observed systems that are outliers from the BHFP relations and {\em not} 
    also outliers from $\mbh-\sigma$ or $\mbh-\mdyn$. 
    \label{fig:outliers}}
\end{figure}

However, given that there {\em is} a systematic dependence on e.g.\ $\mstar$ or $\re$ at fixed 
$\sigma$ which is captured only by the BHFP relations, we expect that the importance of estimating 
BH masses from the BHFP will be enhanced at the extremes of observed distributions. 
Figure~\ref{fig:outliers} compares the locations of outliers in the simulated and observed 
$\mbh-\sigma$ and $\mbh-\mdyn$ relations with their locations on the BHFP relations. 
Indeed, several systems which appear as outliers in one of the projections of the BHFP are no longer 
significant outliers in the BHFP relation. This is true for a number of systems on both the 
$\mbh-\mstar$ relation and 
the $\mbh-\sigma$ relation\footnote{NGC 1023, 3384, 4697, 5252, 
and Cygnus A have 
BH mass measurement errors of $<0.1\,$dex and 
measured masses which are $0.32,\, 0.96,\, 0.81,\, 0.53,\ {\rm and}\ 0.34\,$dex 
discrepant with the expectation from 
the \citet{tremaine:msigma} relation, but only $0.10,\, 0.70,\, 0.51,\, 0.44,\ {\rm and}\ 0.23\,$dex 
discrepant with the BHFP expectation, respectively.}. 
These systems typically have abnormally high or low velocity dispersions given their 
stellar mass, and therefore appear deviant in the BHFP projections, just as such systems 
typically appear to deviate from the spheroid FP in projections 
such as the $\re-\mstar$ or Faber-Jackson ($\mstar-\sigma$) relations. 
Therefore, while the typical scatter about the mean relation is not dramatically different 
for the BHFP ($\sim0.2\,$dex) compared to the $\mbh-\sigma$ relation ($\sim0.3\,$dex), 
the tails of this distribution are substantially suppressed when we adopt the BHFP as a
BH mass estimator. 

\begin{figure}
    \centering
    \figexpand
    \plotter{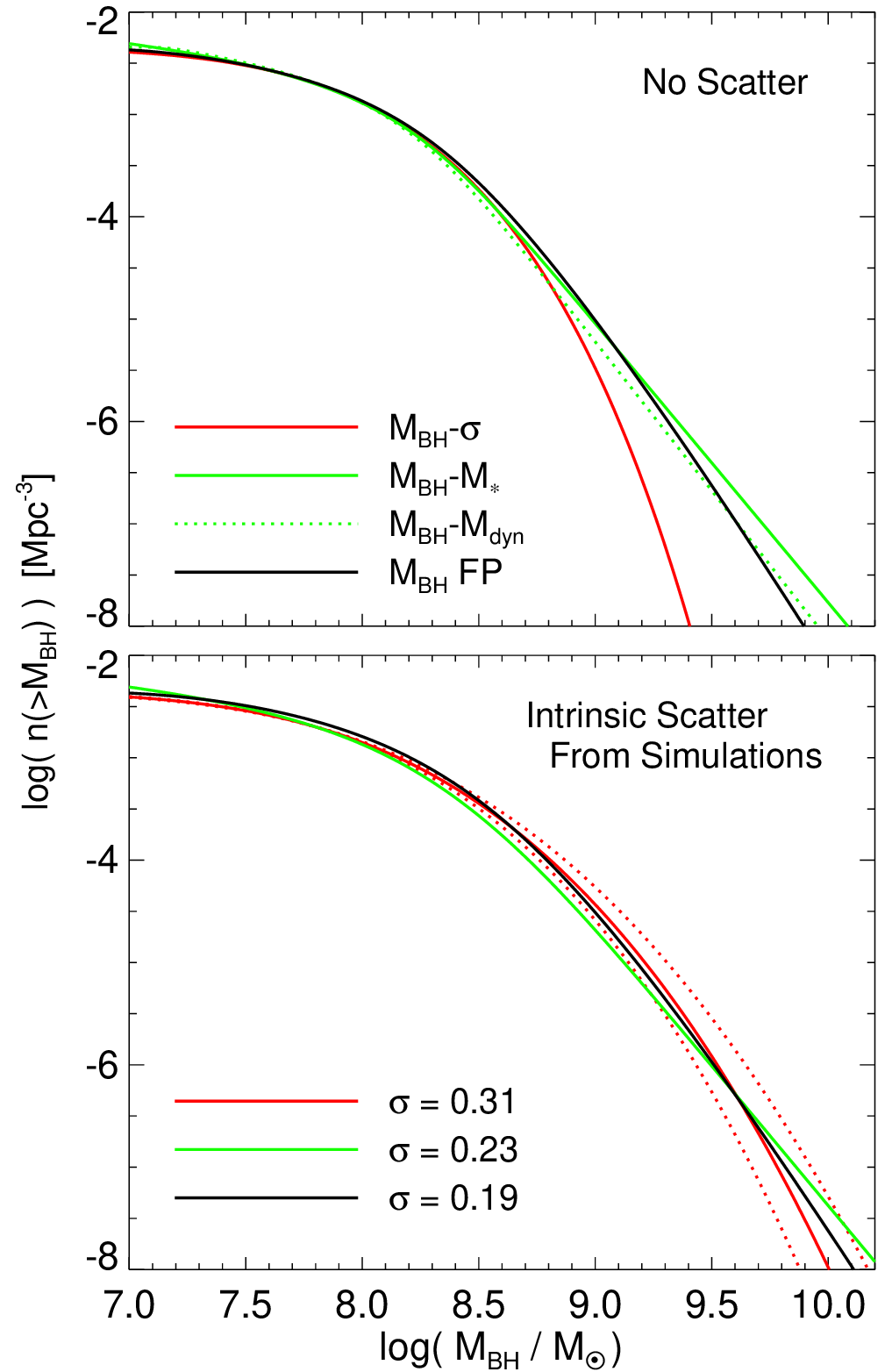}
    \caption{BH mass function (cumulative number density above a given mass) 
    expected from the observed early-type velocity dispersion function 
    (with the $\mbh-\sigma$ relation from 
    \citet{tremaine:msigma}), the observed early-type stellar mass function 
    (with the $\mbh-\mstar$ relation herein, but this is quite similar 
    to the $\mbh-\mdyn$ relation from \citet{haringrix}), and from 
    the joint distribution of $\mstar$ and $\sigma$ (given the observed $\mstar-\sigma$ 
    relations from \citet{bernardi:correlations} or implied by the two distribution functions, 
    and our BHFP relation) or $\sigma$ and $\re$ (with the observed $\mbh-\mdyn$ relation).
    The implied BHMF at high masses is very different 
    for some of the various correlations, if we ignore the scatter in the relations ({\em upper}), with 
    the BHFP representing an intermediate case (albeit closer to $\mbh-\mstar$ than 
    $\mbh-\sigma$). However, accounting for the intrinsic scatter in each correlation 
    estimated from our simulations ({\em lower}) yields nearly identical BHMFs even at 
    $\sim10^{10}\,M_{\sun}$. Larger scatter in the $\mbh-\sigma$ relation at high masses 
    reflects (and compensates for) the discrepancy at high-$\mbh$.
    Dotted lines show the BHMF inferred from $\mbh-\sigma$ 
    with a change of just $25\%$ in the estimated intrinsic scatter (from 
    a scatter of $\sigma=0.31$ to $\sigma=0.27,\ 0.36$), which demonstrates that a 
    small uncertainty in the intrinsic scatter (difficult to determine observationally) 
    can dominate the choice of BH-host correlation adopted, even at very high $\mbh$. 
    \label{fig:bhmf}}
\end{figure}

Given the potential importance of the BHFP for predicting the masses of BHs, especially 
in extreme systems, we should examine the implications that the relation has for 
the local demographics of BHs. There has been substantial debate 
recently about whether or not high-$\mstar$ systems begin to deviate from 
the low-$\mstar$ Faber-Jackson relation \citep[see, e.g.][]{boylankolchin:dry.mergers,
bernardi:magorrian.bias,batcheldor:bcgs}. 
If so, this implies that either the $\mbh-\mstar$ or $\mbh-\sigma$ relation must change 
slope at the highest masses, with one of the two or some other relation being the 
conserved relation. Because the distribution of spheroid velocity dispersions 
\citep{sheth:sigma.distrib} declines more steeply at high $\sigma$ than the galaxy mass or 
luminosity functions do at high $\mstar$ (when BCGs are included), the 
assumption that the $\mbh-\mstar$ relation 
remains unchanged (the simplest expectation from gas-free or ``dry'' mergers) 
naively predicts a much higher abundance of 
very high-mass ($\gtrsim10^{9}\,M_{\sun}$) BHs \citep[for 
an extended discussion, see][]{lauer:massive.bhs}.

Figure~\ref{fig:bhmf} compares the 
expected BH mass function (BHMF) from the observed distribution of 
spheroid velocity dispersions and the $\mbh-\sigma$ relation with that expected 
from the early-type galaxy stellar mass function and the $\mbh-\mstar$ relation. 
We adopt the distribution of velocity dispersions 
$\sigma$ from \citet{sheth:sigma.distrib}, and the 
early-type galaxy stellar mass function from \citet{bell:mfs}, with the 
addition of BCGs at high masses from \citet{lin:bcgs} (yielding a 
shallower power law-like falloff at high masses, as opposed to an 
exponential cutoff). Note that a similar result is obtained 
estimating the BCG mass function from \citet{lauer:massive.bhs} 
or using the \citet{cole:lfs} or \citet{jones:lfs} $K$-band luminosity 
functions from the 2dF and 6dF respectively (converted to 
mass functions following \citet{bell:mfs}), which include BCGs and 
extend to the low space densities of interest. (In any case we are simply 
attempting to highlight the key qualitative behavior, that with the inclusion of 
BCGs the stellar mass function falls off more slowly than the 
velocity dispersion function.)

For now, we assume a one-to-one correlation between $\mbh$ and 
$\sigma$ or $\mstar$. Given this, there would also be a one-to-one correlation between 
$\mstar$ and $\sigma$ (i.e.\ $\sigma(\mstar)$ is given by matching the two 
distributions at fixed number density), so we can use the BHFP relation in the form 
$\mbh\propto \mstar^{\alpha}\,\sigma^{\beta}$ to estimate the BHMF from this relation. 
Unsurprisingly, we find that the BHMF from $\mbh-\sigma$ cuts off rapidly compared 
to that from $\mbh-\mstar$. With a mixed dependence on both $\sigma$ and $\mstar$, 
the BHFP relation predicts a somewhat intermediate case at high masses, although 
it is closer to the expectation from $\mbh-\mstar$.

However, this treatment ignores the important fact that there is scatter in these 
correlations. To predict the BHMF from the distribution of velocity dispersions, we should 
properly convolve over the mean relation broadened by some (approximately lognormal) 
dispersion with width $\sim0.3\,$dex.
The intrinsic scatter is difficult to determine from the observations, if errors 
are not completely understood, so we adopt the intrinsic scatter in each correlation 
estimated from our simulations (Table~\ref{tbl:correlations}). Given this scatter, 
we re-calculate the expected BHMFs, shown in Figure~\ref{fig:bhmf}. Interestingly, 
the three predicted BHMFs are now almost identical, even at very high 
BH masses ($\gtrsim5\times10^{9}$). 
This is expected -- if we completely understand the correlation (and its scatter) between $\mbh$ and 
either $\sigma$ or $\mstar$ over the entire mass range of interest, then any projected 
version of the same BHFP must yield a similar BHMF. 
The fact that the distribution of $\sigma$ cuts off 
more steeply than $\mstar$ is compensated by the fact that the intrinsic scatter 
in $\mbh-\sigma$ is slightly larger than in $\mbh-\mstar$ \citep[see also][]{marconi:bhmf}. 
This is roughly equivalent to the statement that at large masses, 
where the relation between $\sigma$ and $\mstar$ may change, the 
$\mbh-\sigma$ relation must change correspondingly \citep[following the true BHFP 
relation; for a more detailed comparison see][]{lauer:massive.bhs}.
Because the distribution of $\sigma$ falls rapidly at high $\sigma$, 
there is little contribution at low-$\mbh$ from high-$\sigma$ systems, so 
a change in slope or increase in scatter in $\mbh-\sigma$ both have the primary 
effect of increasing the expected number of high-mass BHs, reconciling the BHMFs.

The scatter is of critical importance at these masses: we consider the BHMF derived 
from $\mbh-\sigma$ if we change the estimated intrinsic scatter by just $25\%$ (i.e.\ within 
the range $0.27-0.36$\,dex), all within the range allowed by the present 
observations \citep{tremaine:msigma}, and find that this relatively small difference in the 
intrinsic scatter estimate makes a larger difference at high $\mbh$ than the choice of 
correlation ($\mbh-\sigma$, $\mbh-\mstar$, or BHFP) adopted. 
This reinforces 
the point emphasized by \citet{yutremaine:bhmf,yulu:bhmf,tundo:bhmf.scatter} 
that the estimated intrinsic scatter 
can dominate the demographics of high-mass BHs -- accounting for this, 
the BHFP does not substantially change these estimates.

\section{The Physical Origin of the Fundamental Plane}
\label{sec:origins}

If BH growth terminates because of self-regulation, the 
fundamental requirement is that sufficient energy be released to 
unbind the surrounding galactic gas. 
Given a radiative efficiency $\epsilon_{r}$ and feedback 
coupling efficiency $\eta$, the energy coupled to the IGM from 
accretion onto the BH over its lifetime is 
simply 
\begin{equation}
E_{\rm BH} = \eta\,\epsilon_{r}\,M_{\rm BH}\,c^{2}
\label{eqn:ebh}
\end{equation}
and the binding energy of the gas in the center of the galaxy 
is 
\begin{equation}
E_{\rm gas} = \tilde{\phi}\,\fgas\,\mstar\,\sigma^{2}, 
\end{equation}
where $\tilde{\phi}$ is a constant that depends on the 
shape of the bulge profile ($\tilde{\phi}=10.1$ for a 
\citet{hernquist:profile} profile). In detail, the energy in Equation~\ref{eqn:ebh} 
is proportional to the accreted (as opposed to the total) BH mass, and 
the binding energy of the bulge changes as a function of time during 
the merger. However, this simple estimate 
is actually a reasonable approximation to what occurs in the simulations, 
for two reasons. 

First, as is also demanded by numerous 
empirical constraints \citep[e.g.,][]{soltan,hopkins:bol.qlf}, the majority 
$(\gtrsim70-80\%)$ of the BH growth occurs in the observed 
``quasar'' phase, which is constrained to be both 
short-lived \citep[e.g.][]{martini:lifetimes} and near-Eddington \citep[e.g.][]{kollmeier:mdot}, 
as occurs in the final growth phase near the end of the merger.
Therefore, most of the final $\mbh$ is accreted in the final $e$-folding 
of BH growth, over a Salpeter time $t_{\rm S}\sim4.2\times10^{7}\,{\rm yr}$. 
This is small compared to the cooling time of the 
galactic gas at $\sim\re$, so the approximation that the energy released 
$\sim \mbh$ is reasonable. 
Second, by the time of the quasar phase, the merging galaxies have 
coalesced, and the bulge is largely formed and in place, so the BH growth 
occurs in the relatively fixed potential of the remnant. 

Equating the energy needed to unbind the surrounding gas and 
terminate accretion yields the expected scaling 
\begin{equation}
\mbh \approx 10^{8}\,M_{\sun}\,\frac{0.005}{\eta\,\epsilon_{r}/{f}^{\prime}_{\rm gas}}\,
{\Bigl(}\frac{\mstar}{10^{11}\,M_{\sun}}{\Bigr)}\,
{\Bigl(}\frac{\sigma}{200\,{\rm km\,s^{-1}}}{\Bigr)}^{2}, 
\label{eqn:scaling.expected}
\end{equation}
where we adopt a \citet{hernquist:profile} profile
and a typical $\epsilon_{r}=0.1$, $\eta\sim0.05$ (similar to the values 
adopted in our simulations), which roughly reproduces the observed 
$\mbh-\mstar$ normalization. 
Note that, in detail, the BH feedback need not be radiative -- kinetic 
wind or jet feedback could inject comparable energy 
\citep[e.g.,][]{maraschi:jets,tavecchio:jets,merloni:kinetic.lf}, with some 
efficiency $E = \tilde{\eta}\,M_{\rm BH}\,c^{2}$. This yields an identical 
Equation~(\ref{eqn:scaling.expected}), as  
$\tilde{\eta}$ is functionally equivalent to the previous $\eta\,\epsilon_{r}$ -- 
it does not matter in this derivation whether the net feedback 
efficiency ($\eta\,\epsilon_{r}$) represents a radiative or kinetic mode 
(or some sum of the two), or exactly what fraction of the final BH mass 
is accreted in a given growth phase, as the coupling efficiency 
simply serves to set the normalization of the $\mbh$-host correlations. 
It should also be noted that $\fgas$ changes by a large amount 
during the evolution of typical galaxies or (especially) mergers -- 
the value relevant for Equation~(\ref{eqn:scaling.expected}) is 
some effective value in the central regions near the BH during the 
final $e$-folding(s) of growth (hence the 
notation ${f}^{\prime}_{\rm gas}$), which does {\em not} 
necessarily trace the global or pre-active phase gas 
fraction of the system (see \S~\ref{sec:driving}). 

\begin{figure}
    \centering
    \figexpand
    \plotter{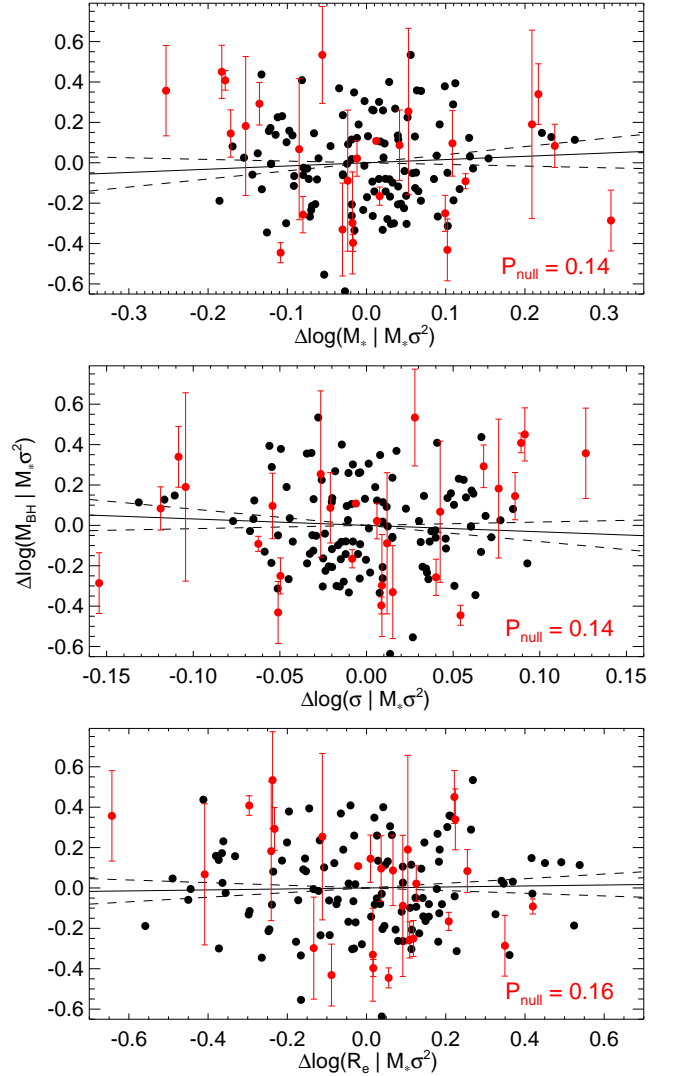}
    \caption{As Figure~\ref{fig:res.reff.sigma}, but comparing the 
    residuals in BH mass and host properties $\mstar$, $\sigma$, and 
    $\re$ at fixed spheroid binding energy $\mstar\,\sigma^{2}$. There 
    is no significant evidence in the simulations or observations for a 
    residual correlation in this space. This is because the fundamental plane in 
    BH mass can be (approximately) represented as a ``tilted'' correlation between BH 
    mass and bulge binding energy, 
    $\mbh\propto (\mstar\,\sigma^{2})^{0.71}$.
    \label{fig:ebinding}}
\end{figure}

We therefore naively expect that the BH mass should scale with $\mstar\,\sigma^{2}$. 
In Figure~\ref{fig:ebinding}, we examine the residuals of the best-fit correlation 
between $\mbh$ and this binding energy $\mstar\,\sigma^{2}$, in the manner of 
Figure~\ref{fig:res.reff.sigma}. In this space, there does not appear to be any strong 
$\gtrsim2\,\sigma$ evidence for a correlation of the residuals in $\mbh(\mstar\,\sigma^{2})$ 
with those of $\mstar$, $\sigma$, or $\re$. 
It seems that the correlation between BH mass and 
bulge binding energy is in some sense more basic
then the correlation between BH mass and e.g.\ $\mstar$ or $\sigma$. 
However, when we fit $\mbh$ to a function of $\mstar\,\sigma^{2}$, we do {\em not} 
recover Equation~(\ref{eqn:scaling.expected}) -- in fact, a linear proportionality 
between $\mbh$ and $\mstar\,\sigma^{2}$ is ruled out at $\sim5\,\sigma$ in 
the observations (and $>10\,\sigma$ in our simulations).  
Instead, BH mass follows a ``tilted'' relation of the form 
\begin{equation}
\mbh\propto (\mstar\,\sigma^{2})^{\alpha}
\end{equation}
with $\alpha\approx 0.71$. Furthermore, this relation is not 
exactly the same as the BHFP we recover from the observations, which is closer 
to $\mbh\propto \mstar^{0.5}\,\sigma^{2}$. 

If we revisit our argument, we note that we have naively assumed 
that the accretion energy from the BH is coupled 
in an (effectively) infinitely short period of time and unbinds the surrounding gas.
More properly, what occurs in our simulations is a pressure-driven 
outflow from the central regions \citep[e.g.,][]{hopkins:faint.slope,hopkins:seyferts}. 
This implies that the necessary condition for self-regulation is the 
injection of sufficient momentum 
\citep[at a rate $\dot{p}\propto L/c$; see e.g.][]{murray:momentum.winds} 
to drive a galactic outflow (total $p\propto \mstar\,\sigma$) within 
the dynamical time 
near the radius of influence of the BH ($R_{\rm BH}\equiv G\mbh/\sigma^{2}$), 
\begin{equation}
t_{\rm dyn}(R_{\rm BH}) \approx \frac{R_{\rm BH}}{\sigma}=\frac{G\,\mbh}{\sigma^{3}}.
\end{equation}
When the rate is below this threshold, it can drive material from the central regions 
where it is initially bound or infalling onto the BH, but the momentum coupled 
is insufficient to entrain the larger-scale material and the outflow fails to halt 
accretion \citep[typical of the early-stage weak winds seen in our simulations 
in earlier merger stages; see Figure~1 in][]{cox:xray.gas}. 
This condition gives us the requirement
\begin{equation}
\dot{p}\,\Delta t\propto \mbh^{2}/\sigma^{3} \propto \mstar\,\sigma, 
\end{equation}
or 
\begin{equation}
\mbh \propto \mstar^{1/2}\,\sigma^{2}, 
\label{eqn:scaling}
\end{equation}
similar to the observed BHFP.

Ultimately, the qualitative conclusions of these derivations are similar. 
Empirical constraints \citep[e.g.,][]{soltan} demand that most BH mass is accumulated  
in high Eddington ratio quasar phases -- so the details of accretion at lower 
Eddington ratios (whether set by e.g.\ the Bondi-Hoyle rate or other accretion 
mechanisms) are not important. This also implies that the BH mass is accumulated in a 
short period of time $\sim10^{7}-10^{8}$ years, such that whether growth is 
driven at 
the end of galaxy mergers, or in secular disk or bar instabilities, the 
environment (background potential) local to the BH is relatively fixed. The BH 
then self-regulates when it is sufficiently massive that its 
feedback energy can unbind infalling gas and halt accretion. Effects which 
deepen 
the local potential at the galactic center and increase the binding energy of 
gas near the BH will prevent gas from being unbound until the BH has grown 
more massive than it would otherwise. For example, as we show in detail in 
\S~\ref{sec:driving}, if spheroid progenitors are more gas-rich, there is more gas 
which can cool and form stars on small scales in the center of the spheroid, 
deepening the potential well there (i.e.\ 
yielding a more concentrated remnant with smaller $\re$ and larger $\sigma$, 
while having little effect on the total stellar mass). This requires that any BH grow 
larger in order to unbind nearby gas and halt accretion, in a manner 
similar to the scaling of Equation~(\ref{eqn:scaling}). 

We might also ask how sensitively this BHFP scaling or ``tilt'' is 
related to dissipational processes, as 
\citet{robertson:fp} demonstrate that the origin of the spheroid FP tilt lies 
essentially in the scale-dependence of dissipational processes such as 
gas cooling and star formation. From the derivation above, we would 
expect that although the processes from \citet{robertson:fp} might affect the structure 
of the merger remnants themselves, they should not change how, fundamentally, the central 
BH self-regulates. Unfortunately, BH accretion is, itself, naturally
a dissipational process, and therefore we cannot simply test this theory 
in our case by running simulations where gas dissipation is turned off.

However, \citet{robertson:fp} further show that it actually requires substantial 
initial gas fractions for these dissipational effects to act -- in mergers with very 
low initial gas fractions ($\lesssim20\%$) no ``tilt'' (in the spheroid FP) is induced,
consistent with requirements from the observed phase space densities of
ellipticals \citep[e.g.,][]{hernquist:phasespace}.
We therefore 
briefly consider just a set of simulations with initial $\fgas=0.05$. This is sufficiently low 
that the remnants act dissipationlessly, and lie on the virial relation as opposed to 
the spheroid FP (see Figure~10 of \citet{robertson:fp}), but sufficiently large that we 
do not need to worry about artificially ``strangling'' the BH by giving it insufficient 
gas to accrete (given typical $\mbh\approx 0.001\,\mstar$, the BH need only 
have access to $\sim2\%$ of this gas to grow normally). 

We find that these merger remnants obey a similar BHFP relation to their 
high-$\fgas$ counterparts, implying that so long as feedback-driven self-regulation (as opposed to 
e.g.\ gas starvation) determines the final BH mass, these scalings are robust. 
However, the significance of the preference for e.g.\ a BHFP relation as opposed to 
a simpler $\mbh\propto\mstar$ or $\mbh\propto\sigma^{4}$ relation is greatly reduced. 
This is because, without significant effects of dissipation to change the central phase-space 
structure and potential depth of the remnant, the velocity dispersion $\sigma$ and effective 
radius $\re$ are simply set by the violent relaxation of the scattered stellar disks. 
The general scalings of the BHFP are not, then, unique to gas-rich progenitors, but 
their significance and the importance of accounting for the observed dependencies are so. 

Indeed, regardless of prescriptions for BH accretion, feedback, 
and star formation, the above derivations (and qualitative dependence on 
galaxy properties) depend on only three relatively robust assumptions -- 
that BH growth is dominated by bright, high-Eddington ratio phases, 
that feedback from accretion affects the gas on small scales around the BH 
(sensitive to the central potential of the galaxy), 
and that some sort of heating, momentum coupling, or effective pressure eventually halts 
accretion. This does place some constraints on scenarios for BH growth, however. 
In models where, for example, the BH accretion rate 
is a pure function of the galactic star formation rate 
\citep[see e.g.,][although in some of these 
cases the Eddington limit is still preserved as 
an upper limit]{kawakatu,monaco,granato,cattaneo}, it is difficult to explain how 
the BH mass would be sensitive to the central potential in 
the manner of the observed BHFP, and not simply trace the galaxy 
stellar mass (i.e.\ a yield pure $\mbh-\mstar$ relation, which the observations 
disfavor at $\sim3\,\sigma$).

\section{Driving Systems Along the Fundamental Plane}
\label{sec:driving}

We have thus far considered systems in terms of the observable properties of 
the remnants, specifically quantities like $\mstar$, $\sigma$, and $\re$. 
We now turn to the ``theorist's question'' -- namely, how are the 
positions of systems on the $\mbh-\mstar$, $\mbh-\sigma$, and BHFP relations 
affected by theoretical quantities or initial conditions? 

\begin{figure}
    \centering
    \figexpand
    \plotone{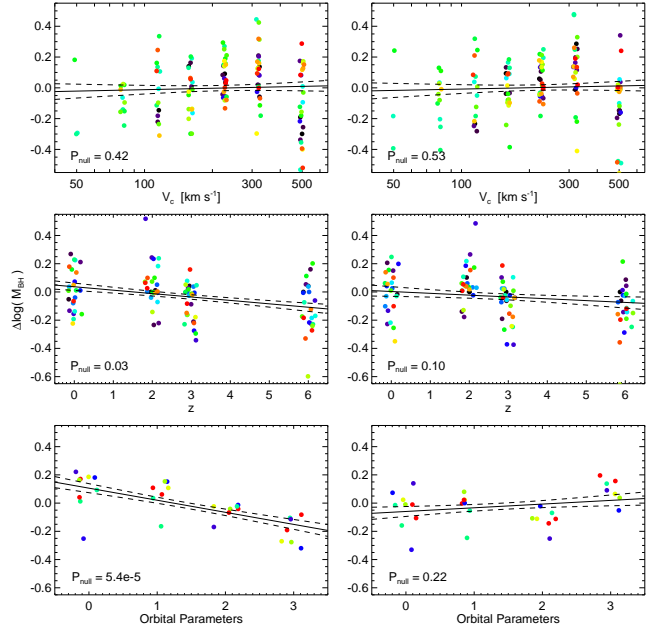}
    \caption{Residuals of the $\mbh-\mstar$ ({\em left}) 
    and BHFP ({\em right}) relations as a function of 
    various initial conditions or theoretical quantities. $V_{c}$ is the 
    halo virial velocity, $z$ is the redshift to which the progenitor disks 
    are initialized (i.e.\ their concentrations and scale lengths are rescaled by a 
    small amount to match disks at these redshifts), and 
    the orbital parameters shown span a small set ranked from those that yield the 
    least (0) to most (3) rotation in the merger remnants \citep[average 
    $(V_{c}/\sigma)^{\ast}=0.06,\,0.25,\,0.41,\,1.03$, respectively; for details, see][]{cox:kinematics}. 
    By changing 
    e.g.\ the circular velocities, initial disk formation redshifts, or orbital 
    parameters of simulated mergers, we can drive changes in the 
    $\mbh-\mstar$ relation. However, these can all be understood 
    in terms of how they change the galaxy structure -- i.e.\ they drive 
    changes in $\sigma$ at fixed $\mstar$, while preserving the 
    BHFP relation. Higher angular momentum mergers also produce more disk-like 
    remnants -- but when we 
    are careful to exclude the contribution from 
    rotation to the velocity dispersion $\sigma$, the objects lie on the expected BHFP.
    \label{fig:residuals.initialconditions}}
\end{figure}

Figure~\ref{fig:residuals.initialconditions} considers the residuals in 
the $\mbh-\mstar$ and BHFP relation for merger remnants with different 
virial velocities, redshifts, and orbital parameters. 
Note that ``redshift'' in this context simply refers to the characteristics of 
the progenitor disks (which are initialized to resemble
disks at low redshifts $z\sim0$, or higher, $z=2,\,3,\,6$). 
We select each set of simulations 
in which all parameters are identical except that plotted in the 
figure, then consider the residual 
with respect to the mean relation for just those simulations (each 
set of points of different color plotted in the figure represents one such 
set of simulations). 

In terms of $\mbh/\mstar$, there are weak 
($\sim0.2-0.3$\,dex over the maximal 
range spanned by the simulations) trends towards lower $\mbh/\mstar$ for higher 
redshift and larger angular momentum mergers, but the systems lie on the 
same BHFP regardless of $\vvir$, $z$, orbital parameters, or any other initial quantities
we vary. 
The trends in $\mbh/\mstar$ simply 
reflect changes in the structural properties of the 
remnants -- for example, particular orbital parameters produce 
smaller bulges and lower values of $\sigma$ at fixed $\mstar$. This is 
expected -- because 
most of the BH growth occurs over a short period of time at 
the end of a merger or other phase of activity (the last 1-2 $e$-folding times) 
during which the environment is relatively dynamically settled, 
changing initial conditions should only affect $\mbh$ indirectly by 
altering the central structure of the remnant. 
The lack of a strong trend in $\vvir$ simply reflects the 
fact that the central regions which set the potential depth of 
relevance for determining $\mbh$ are very much baryon-dominated. 
This is not to say that $\mbh$ does not scale with $\vvir$, in a mean sense, 
but simply that the dependence on $\vvir$ is subsumed in the more 
direct dependence on $\mstar$ and $\sigma$, which themselves can 
depend on halo mass or $\vvir$. 

For this reason, at fixed $\mstar^{1/2}\,\sigma^{2}$, we expect BH mass to 
be independent of halo mass and redshift. Of course, $\mstar$ and $\sigma$ will, 
in the mean, scale with $\mhalo$, which could yield evolution in the 
$\mbh-\mhalo$ relation. For a bulge-dominated system (i.e.\ ignoring the complication 
that the same mass halo could host a disk-dominated system with 
a much smaller BH), $\mstar = f_{\ast}\,(\Omega_{b}/\Omega_{m})\,\mhalo$, where 
$f_{\ast}(\mhalo,\,z)$ is the typical fraction of baryons incorporated into the galaxy. 
In the simulations, we find a rough correlation $\sigma\propto v_{\rm max}$ 
(similarly, $v_{\rm c}\propto v_{\rm max}$ for our progenitor disks, although that 
is by construction), 
where $v_{\rm max}$ is the maximum halo circular velocity 
($\propto\vvir\,c^{1/2}$, where $c(\mhalo,\,z)$ is the halo concentration), 
modulo the effects of e.g.\ gas fraction and orbital parameters changing $\sigma$ 
at fixed $\mhalo$. Observationally, a similar mean correlation 
\citep[$\sigma\approx 0.66\,v_{\rm max}$, nearly identical to the 
best-fit normalization in our simulations; e.g.][]{kronawitter:ell.dynamics,gerhard:ell.dynamics} is found. 
Given $c\propto \mhalo^{-0.13}\,(1+z)^{-1}$ \citep{bullock:concentrations}, this implies 
$\mbh \propto \alpha(z)\,f_{\ast}^{0.5}\,\mhalo^{1.04}$, where $\alpha(z)$ 
represents the weak remaining redshift evolution term, 
$\alpha \equiv (\Omega_{m}\,\Delta_{c}(z) / \Omega_{m}(z)\,18\,\pi^{2})^{1/3}$, 
which changes by only $\sim20\%$ from $z=0-6$. We therefore expect that 
evolution in the $\mbh-\mhalo$ relation will be dominated by 
the effects of evolution in typical gas fractions and remnant structural 
properties on $\sigma$ (changing $\sigma$ at fixed $\mhalo$ in a systematic 
sense) as well 
as cosmological evolution in typical baryon incorporation fractions 
$f_{\ast}$ and/or bulge-to-disk ratios in galaxies hosted by 
halos of a given mass. To the extent that such effects occur, they of course 
should be also traced in  
some mean evolution in e.g.\ the $\sigma-v_{\rm max}$ relations.

%The lack of a strong trend in $\vvir$ is contrary to the naive expectation if, for example, the bulge 
%velocity dispersion $\sigma$ simply traced $\vvir$ -- reflecting the 
%fact that the central regions which set the potential depth of 
%relevance for determining $\mbh$ are very much baryon-dominated. The 
%(weak) trend with redshift simply reflects changes in the structural properties of 
%the progenitors, as shown in \citet{robertson:msigma.evolution} -- these structural changes 
%slightly shift the typical $\sigma(\mstar)$, yielding some evolution in $\mbh/\mstar$. 
%The trend in orbital parameters is easily understood, 
%as particular (e.g.\ polar or high-angular momentum) encounters 
%yield diskier, more rotationally-supported remnants that have 
%smaller bulges and intrinsic $\sigma$ values, thus smaller BHs for the same $\mstar$.
%At a given stellar mass, then, it is possible to drive relatively weak evolution 
%in $\mbh/\mstar$ by changing either the structural properties or orbits of 
%the progenitors. However, the differences are not dramatic (typically within 
%observational systematic uncertainties). Moreover, the remnants still lie 
%on the same BHFP -- the deviation with respect to $\mbh/\mstar$ derives from 
%driving structural changes in the remnant, i.e.\ changing $\sigma$ at fixed $\mstar$, 
%but there is no trend in the residuals with respect to the BHFP with 
%$\vvir$, $z$, orbital parameters, or any other quantities we vary. 

We have also studied a large subset of simulations of different masses 
and gas fractions with a variety of prescriptions for feedback and winds 
from star formation. These results will be discussed in 
detail in Cox et al.\ 2007 (in preparation), but we 
briefly summarize their relevant conclusions. 
The inclusion of massive stellar winds (with 
high mass loading efficiencies $\eta_{\rm w}\gtrsim1$, where 
$\dot{M}_{\rm wind}=\eta_{\rm w}\,\dot{M}_{\ast}$) can affect the 
structure of merger remnants, although not necessarily in a 
monotonic or easily predictable fashion. For example, 
strong winds can remove gas from the central regions of the 
galaxy and yield a lower effective gas fraction $\fgas$, but they 
can also cycle such gas at earlier stages, preventing it from 
immediately being turned into stars before the final merger, 
and actually raising the effective $\fgas$ at the final coalescence. In 
any case, the effects are usually small (especially in the most 
massive galaxies of interest here) or comparable to those owing 
to the choice of orbital parameters. 

More important, regardless of 
the stellar feedback efficiency, star formation alone cannot prevent gas 
from accreting onto the central BH (especially given that only $\sim0.1\%$ of 
the galaxy mass in gas needs to reach the BH to affect the 
$\mbh$-host correlations), and in fact strong stellar winds can 
(either by cycling gas as above, or by shocking and increasing 
the supply of low angular momentum material) greatly increase 
the fuel supply for accretion at the galactic center. The correlation 
between BH mass and host properties is therefore, regardless of the prescription for 
stellar feedback, still set by the local self-regulation of the BH, and 
obeys (in all our simulations) an identical BHFP relation. Stellar winds 
can, in principle, influence the final BH mass, but only indirectly 
by affecting the structure of the remnant, in the same manner 
as changing progenitor structural properties, orbital 
parameters, and gas fractions.

\begin{figure}
    \centering
    \figexpand
    \plotter{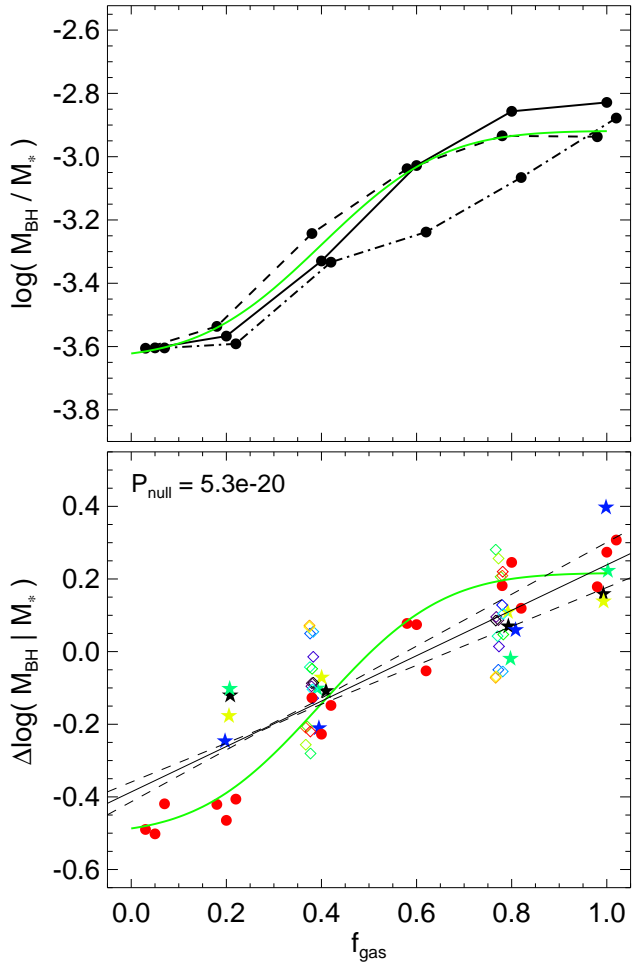}
    \caption{Trend in $\mbh/\mstar$ as a function of simulation initial 
    disk gas fraction.
    Upper panel plots $\mbh/\mstar$ for three suites of simulations 
    (solid, dashed, and dot-dashed lines). Each of the three is 
    a set of otherwise identical mergers of Milky Way-like systems, varying only the 
    gas fraction with values $\fgas=0.05,\ 0.2,\ 0.4,\ 0.6,\ 0.8,\ 1.0$. The three 
    suites consider three different orbital configurations for the merger. 
    Lower panel considers these residuals and those from our other simulations 
    versus $\fgas$ as in Figure~\ref{fig:residuals.initialconditions}. 
    Black line is a log-linear fit (dashed show $\pm1\,\sigma$), green 
    line the more accurate fit $\log{(\mbh/\mstar)}=-3.27+0.36\,{\rm erf}{[(\fgas-0.4)/0.28]}$. 
    High-$\fgas$ mergers produce much larger BHs than low-$\fgas$ systems. 
    \label{fig:residuals.fgas}}
\end{figure}

Figure~\ref{fig:residuals.fgas} considers the trend in $\mbh/\mstar$ as a function of 
the initial gas fraction of our simulations. In contrast to trends with 
$\vvir$, orbital parameters, or the evolution of disk structural parameters with redshift, 
the dependence of $\mbh/\mstar$ on $\fgas$ is quite strong, varying by 
nearly an order of magnitude from low ($\fgas\lesssim0.2$) to high 
($\fgas\gtrsim0.8$) initial gas fractions. We consider this in detail by examining 
a small case study set of simulations. We construct a fiducial set of simulations of 
Milky-Way like initial disks ($\vvir=160\,{\rm km\,s^{-1}}$), and collide them in 
otherwise identical mergers except for varying the initial disk gas fractions 
with values $\fgas=0.05,\ 0.2,\ 0.4,\ 0.6,\ 0.8,\ 1.0$. We construct three such 
suites, each with a different orbit (roughly bracketing the extremes of 
possible merger configurations). Figure~\ref{fig:residuals.fgas} also shows 
the trend of $\mbh/\mstar$ in these simulations -- it is clear that, all else being 
equal, larger values of $\fgas$ drive the systems to larger 
$\mbh/\mstar$. 

\begin{figure*}
    \centering
    \figexpand
    \plotone{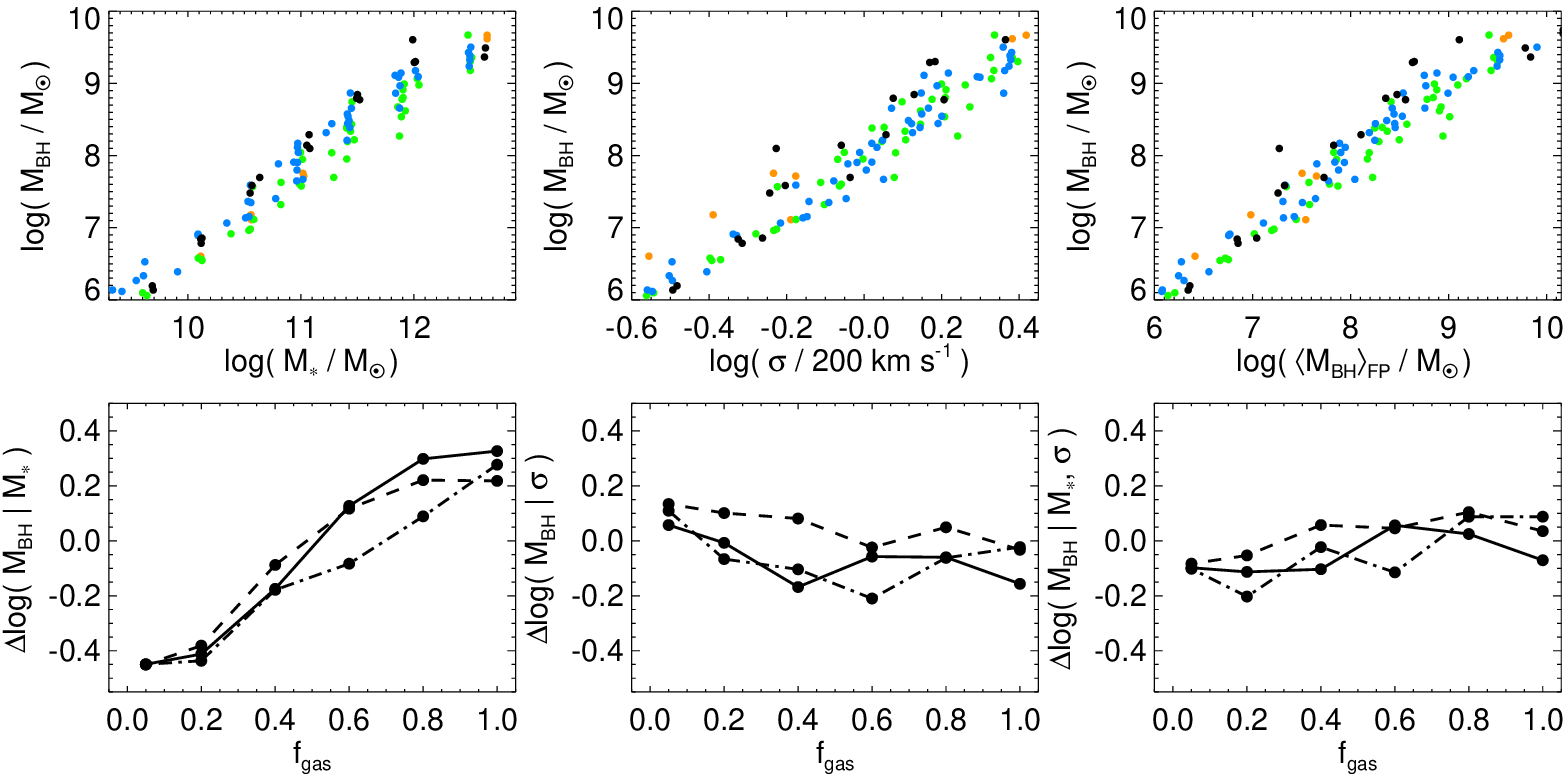}
    \caption{{\em Left:} Correlation of $\mbh$ with $\mstar$ in simulations 
    with different gas fractions $\fgas=0.2,\ 0.4,\ 0.8,\ 1.0$ (orange, green, 
    blue, and black, respectively; {\em upper}), and residuals of $\mbh-\mstar$ as a 
    function of $\fgas$ for our case study simulations from Figure~\ref{fig:residuals.fgas} ({\em lower}). 
    {\em Center:} Same, but for the $\mbh-\sigma$ correlation. 
    {\em Right:} Same, but for the BHFP relation. 
    Varying $\fgas$ does not move the remnants off the BHFP relation, but does 
    systematically shift them with respect to $\mstar$ and $\sigma$. 
    The trend of 
    $\mbh/\mstar$ with $\fgas$ in Figure~\ref{fig:residuals.fgas} 
    must therefore fundamentally relate to how $\fgas$ modifies 
    structural properties like $\sigma$ (preserving the BHFP), 
    not to the naive expectation that larger $\fgas$ translates to 
    ``more material to be unbound.''    
    \label{fig:mbh.vs.msigma2}}
\end{figure*}

However, the trend here does {\em not} resemble the 
simple $\mbh\propto \fgas$ scaling that we naively predicted in 
Equation~(\ref{eqn:scaling.expected}) by demanding that the BH be able to 
unbind the entire initial gas content of the galaxy. In fact, Figure~\ref{fig:mbh.vs.msigma2} 
shows the correlation between $\mbh$ and $\mstar\,\sigma^{2}$ for simulations 
of different gas fractions, and there is no systematic trend with $\fgas$. Likewise, the 
remnants lie on the BHFP regardless of their gas fractions -- i.e.\ the change 
in $\mbh/\mstar$ can be entirely accounted for by the change in $\sigma$ at 
fixed $\mstar$. 
This should, perhaps, not 
be surprising -- our earlier derivation neglected the fact that, by the 
quasar phase and epoch of final BH growth, the large majority 
($\gtrsim80-90\%$) of $\fgas$ has already been turned into stars or ejected 
by stellar winds. The trend in $\mbh/\mstar$ with $\fgas$ must therefore be 
primarily driven by how $\fgas$ changes the structural properties of the 
remnant, a more subtle effect than the naive ``amount of material to be unbound'' expectation. 

\begin{figure}
    \centering
    \figexpand
    \plotone{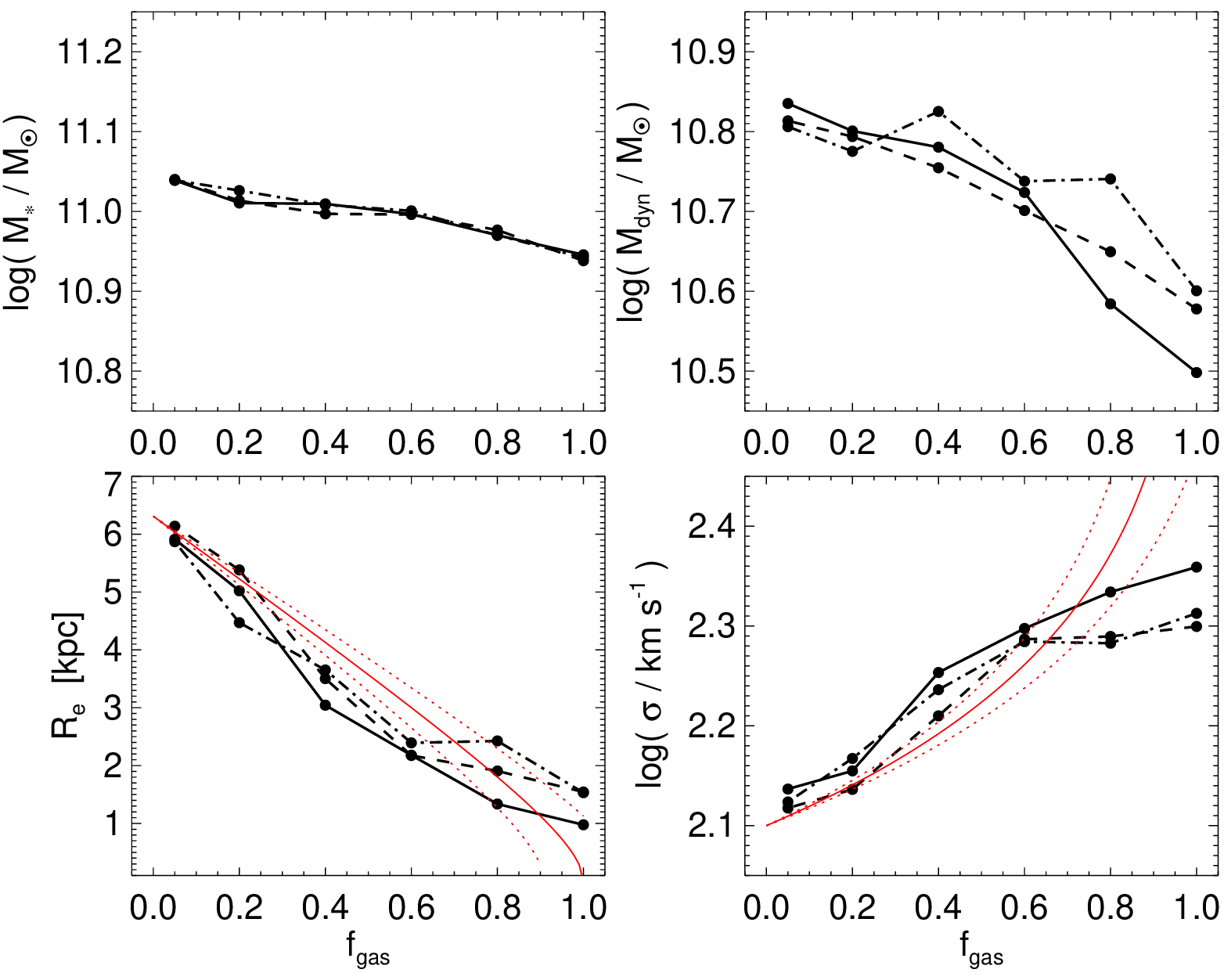}
    \caption{The structural properties of the merger remnants from 
    our set of gas fraction case studies from Figure~\ref{fig:residuals.fgas}. 
    Changing $\fgas$ has almost no effect on $\mstar$, but 
    by increasing the amount of dissipation and fraction of stellar material formed 
    in the final, central starburst, increasing $\fgas$ produces more 
    concentrated remnants with smaller effective radii $\re$ and larger 
    central velocity dispersions $\sigma$. Red lines show the expectation of 
    a toy model in which a fraction $\sim0.5$ (solid; $\pm0.05$, dashed) 
    of the gas participates in a central 
    starburst (Equation~\ref{eqn:fgas.reff}).
    \label{fig:fgas.casestudy}}
\end{figure}

In order to understand this strong dependence of $\mbh/\mstar$ on $\fgas$, we 
consider the structural properties of the merger remnants in our case study 
in Figure~\ref{fig:fgas.casestudy}. On inspection, it is clear that the trend in 
$\mbh/\mstar$ is one of increasing $\mbh$ at fixed $\mstar$, as 
$\mstar$ is nearly constant with $\fgas$ (there is a weak 
trend, as not all of $\fgas$ is converted to stars, but this changes $\mstar$ 
by $<0.1$\,dex from $\fgas=0$ to $\fgas=1$). However, there is a strong 
trend in $\re$ and $\sigma$ with $\fgas$, as increasing the amount of 
dissipation (through $\fgas$) yields more concentrated remnants. 

\begin{figure*}
    \centering
    \figexpand
    \plotone{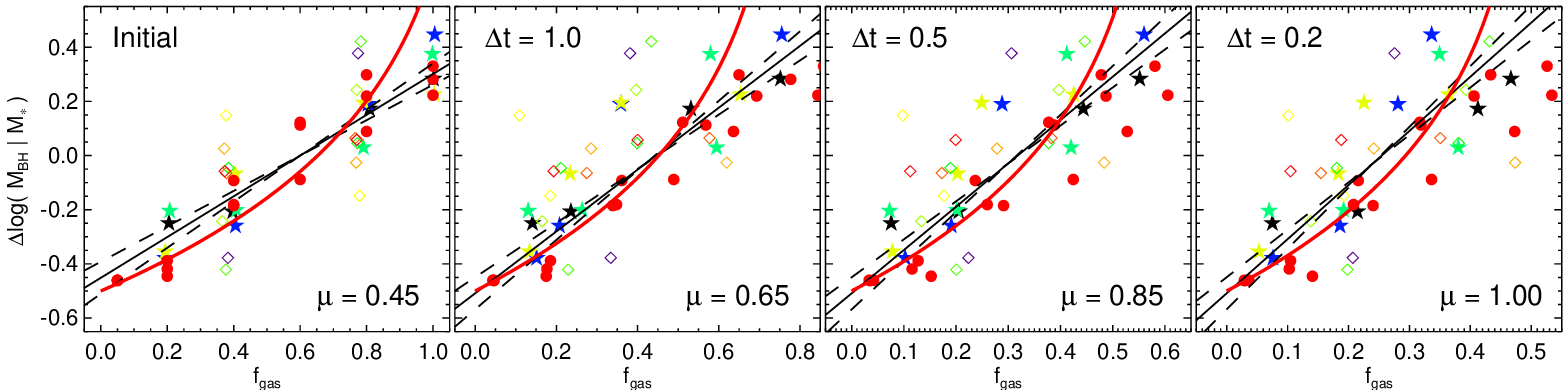}
    \caption{Dependence of $\mbh/\mstar$ on $\fgas$, as in Figure~\ref{fig:residuals.fgas}. 
    Panels plot this as a function of $\fgas$ measured at the initial time in each simulation 
    ({\em left}) and at times $\Delta\,t=1.0,\ 0.5,\ {\rm and}\ 0.2$\,Gyr before the 
    final merger and quasar phase. Black lines in each plot the 
    best-fit trend (solid; with $\pm1\sigma$ range in dashed lines), and red lines 
    plot the expectation from our simple model (Equation~\ref{eqn:fgas.reff}), assuming 
    that the BHFP is always preserved and that a fraction $\mu$ of the gas mass 
    measured at each time will participate in the final, central starburst ($\mu$ is smaller 
    at early times because much of this gas will form stars in the two disks well before 
    they merge). Regardless of when $\fgas$ is defined, the trend is similar, 
    and evolution in $\mbh/\mstar$ is driven by preserving the BHFP and increasing 
    $\sigma$ at fixed $\mstar$ in the same manner. 
    \label{fig:fgas.vstime}}
\end{figure*}

We can 
understand this behavior with a simple toy model. Assume that the stars 
formed in the disk(s) before and during the merger (i.e.\ those that will 
act dissipationlessly) 
are scattered into a typical bulge with a \citet{hernquist:profile} profile 
with scale length $\re(\fgas=0)$, independent of the central gas content, and that a 
fraction $\mu$ of the initial gas mass ($M_{\rm gas} = \fgas\,M_{\rm gal}$) 
survives to the late stages of the merger 
and, via dissipation, falls to the center of the galactic potential. There, it 
will form a highly concentrated 
central stellar component (scale length $\ll \re(\fgas=0)$; 
for simplicity we take it to be effectively a point concentration). 
The total enclosed mass as a function of radius is then 
\begin{equation}
M_{\rm gal}(<r) = M_{\rm gal}\,{\Bigl[}\mu\,\fgas + 
\frac{(1-\mu\,\fgas)\,r^{2}}{(r + \re[\fgas=0])^{2}}{\Bigr]}.
\end{equation}
This yields a 
half-mass effective radius $\re$ of 
\begin{eqnarray}
\label{eqn:fgas.reff}
 \frac{\re}{\re(\fgas=0)} & =  &\frac{x}{(\sqrt{2}+1)\,(1-x)}  \\ 
 \nonumber x & =  &{\Bigl(}\frac{1/2-\mu\,\fgas}{1-\mu\,\fgas}{\Bigr)}^{1/2}.
\end{eqnarray}
Figure~\ref{fig:fgas.casestudy} compares this simple expectation for 
$\re$ and $\sigma^{2}\propto\mstar/\re$ 
with that from the simulations as a function of $\fgas$ -- for a 
representative $\mu=0.5$, our toy model describes the simulations 
quite well (until $\fgas\rightarrow1$, where our assumption that the 
inner stellar component is infinitely concentrated breaks down; this 
``saturation'' in $\sigma$ reflects the fact that even extremely gas-rich systems 
will still turn much of that gas into stars that scatter into some large orbits, and may be important for 
e.g.\ the steep observed cutoff in the observed $\sigma$ distribution). 
Note that this scaling is not very sensitive to our assumption about the exact profile shape -- 
assuming the bulge follows an exact $r^{1/4}$ law or adopting a different 
inner power-law slope changes the predicted scaling by only $\sim10-20\%$.
Given this change in $\re$ at fixed $\mstar$ owing to increasing 
$\fgas$, the fundamental plane implies that $\mbh$ should increase, 
roughly as $\sim \re^{-1}$.
Figure~\ref{fig:fgas.vstime} plots the dependence of $\mbh/\mstar$ on 
$\fgas$, compared with the expectation from this simple model. From 
the agreement here, and the fact that Figure~\ref{fig:fgas.casestudy} finds 
no change in the BHFP with $\fgas$, we conclude that more gas-rich mergers drive 
evolution in $\mbh$ by producing more concentrated remnants with 
smaller $\re$ and larger $\sigma$ at fixed $\mstar$, and therefore 
larger $\mstar\sigma^{2}$. 

There is one important caveat to our discussion of disk gas fractions $\fgas$. 
Lacking a full cosmological simulation in which to determine how 
gas continuously accretes onto the disks, we have simply referred to $\fgas$ as 
the initial gas fraction in our simulations. Of course, during a 
simulation, $\fgas$ will decrease as gas is turned into stars, so that the 
actual gas fractions by the time the systems merge may be substantially lower than 
the numbers we quote. In Figure~\ref{fig:fgas.vstime}, 
we reproduce our plot from Figure~\ref{fig:residuals.fgas} of the residuals 
in $\mbh/\mstar$ as a function of the initial gas fraction of the simulations. However, 
we also return to the simulations and measure the gas fraction for each at a set of 
uniform times $\Delta\,t=1.0,\ 0.5,\ {\rm and}\ 0.2$\,Gyr before the final merger 
(defined for convenience as the coalescence of the two BHs). As the 
simulations approach the final merger event, it is clear that the trend of residuals 
with $\fgas$ is qualitatively unchanged -- 
however, the absolute values of $\fgas$ systematically decrease. By $\Delta\,t=0.2\,$Gyr, 
the gas fractions are systematically lower by a factor $\sim2-3$. Therefore, the 
exact values of $\fgas$ which we quote should not be taken too literally -- if 
gas is accreted in the real universe such that $\fgas$ changes less rapidly in earlier 
stages of a merger, then the initial gas fraction need only be as large as $\fgas\sim0.3$ to 
be equivalent to our most extreme $\fgas=1$ cases (with our more typical 
$\fgas=0.4-0.5$ cases corresponding to rather moderate pre-merger or $\Delta\,t=0.2$\,Gyr 
gas fractions of $\sim0.2$).

\section{Implications for the Redshift Evolution in BH-Host Relations}
\label{sec:evolution}

\subsection{Empirical Predictions}
\label{sec:evolution:empirical}

Given that the BHFP appears to be robust against all varied quantities in 
our simulations, we expect that it should be preserved at all redshifts. 
However, this implies that, at fixed $\mstar$, evolution with redshift in the typical 
velocity dispersions and/or effective radii of spheroids 
will also manifest as evolution in the typical $\mbh/\mstar$ relation. 

\begin{figure*}
    \centering
    \plotone{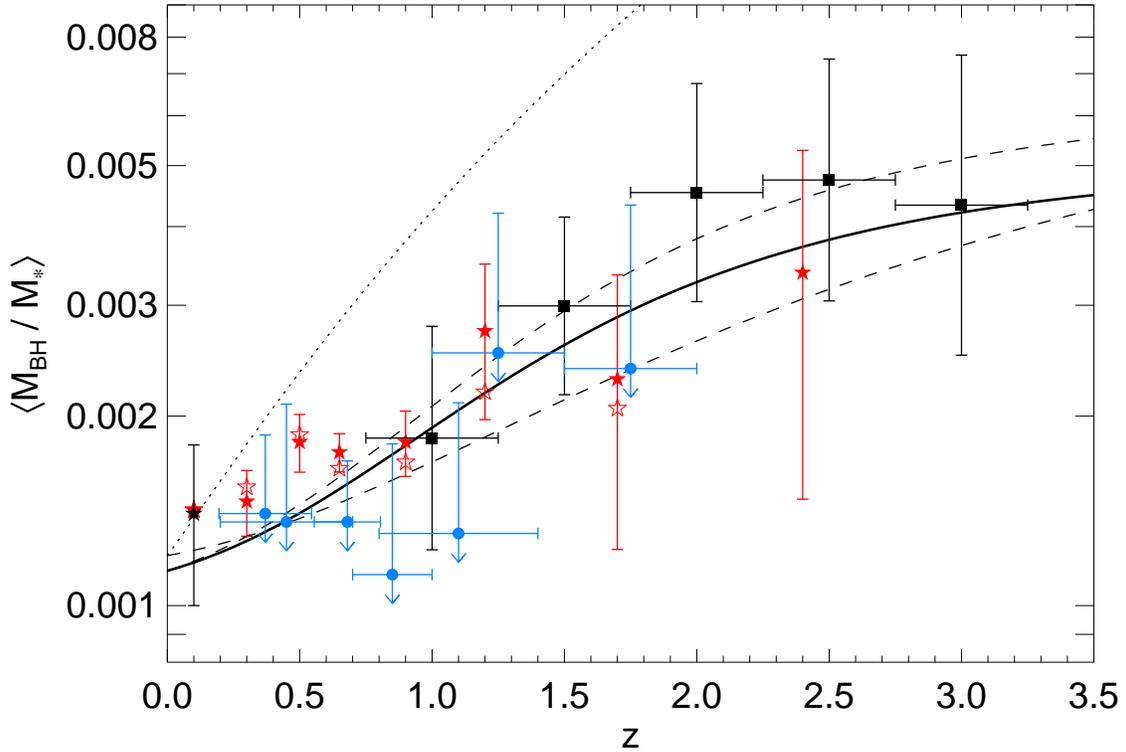}
    \caption{Evolution in $\mbh/\mstar$ as a function of redshift. Red stars 
    plot the evolution expected at fixed $\mstar$ 
    (filled $\sim 6\times10^{10}\,M_{\sun}$; open $\sim3\times10^{10}\,M_{\sun}$) given the 
    observed evolution in the effective radii $\re(\mstar)$ for systems of this 
    stellar mass from \citet{trujillo:size.evolution} and our fundamental 
    plane relations (Table~\ref{tbl:correlations}). Other points compare 
    observational estimates of this evolution: black squares are direct measurements 
    of high-redshift BH masses and host luminosities \citep{peng:magorrian.evolution}, blue circles 
    are upper limit estimates from \citet{hopkins:msigma.limit} based on observed 
    spheroid and BH mass functions. All points are normalized to the same local 
    value (black star) from \citet{haringrix}.
    Solid and dashed lines show the predicted evolution 
    from an {\em a priori} model in which the evolution in $\re(\mstar)$ (and, correspondingly, 
    $\mbh(\mstar)$) is driven by increasing disk gas fractions with redshift, 
    given the best-fit scalings of $\mbh(\fgas\,|\,\mstar)$ (solid; dashed show $\pm1\,\sigma$). 
    The BHFP predicts that as high-redshift spheroids are more compact, $\mbh/\mstar$ must 
    rise, in good agreement with the observations; 
    the trend is driven by more gas-rich progenitors in spheroid-producing mergers. 
    To contrast, dotted line shows the expectation of the simplified (but common) semi-analytic 
    assumption that $\mbh\propto \sigma^{4}\propto V_{c}^{4}$ (assuming 
    $\mstar(M_{\rm vir})$ does not change much with $z$ for $\sim L_{\ast}$ galaxies). 
    \label{fig:empirical.trujillo}}
\end{figure*}

It is observed empirically \citep{trujillo:size.evolution} 
and expected theoretically \citep{khochfar:size.evolution.model} that 
high-redshift spheroids will be more compact 
at a given stellar mass $\mstar$ than their low-redshift analogues. 
Specifically, \citet{trujillo:size.evolution} compile a number of 
measurements of the evolution, relative 
to $z=0$, of the effective radii of spheroids (defined as systems 
with Sersic indices $\sersic>2.5$) at fixed 
stellar mass $\mstar > 3\times10^{10}\,M_{\sun}$ and 
$>6\times10^{10}\,M_{\sun}$, corresponding to typical $L_{\ast}$ 
galaxies at most redshifts. If the BHFP is preserved, this 
necessarily implies evolution in the $\mbh-\mstar$ relation, of the 
form $\mbh/\mstar \propto \re(\mstar)^{-1}$ (see Table~\ref{tbl:correlations} for exact values). 

Figure~\ref{fig:empirical.trujillo} plots this expected evolution 
in $\mbh/\mstar$ from the \citet{trujillo:size.evolution} measurements, normalized to 
the value observed at the same stellar mass by \citet{haringrix} at $z\approx0$. 
For comparison, we plot the estimated evolution in $\mbh/\mstar$ from 
\citet{peng:magorrian.evolution} (specifically, we adopt their early-type template to convert 
their measured luminosities to stellar masses), 
and limits on this evolution from \citet{hopkins:msigma.limit}, 
normalized to the same local value. 

\begin{figure}
    \centering
    \figexpand
    \plotone{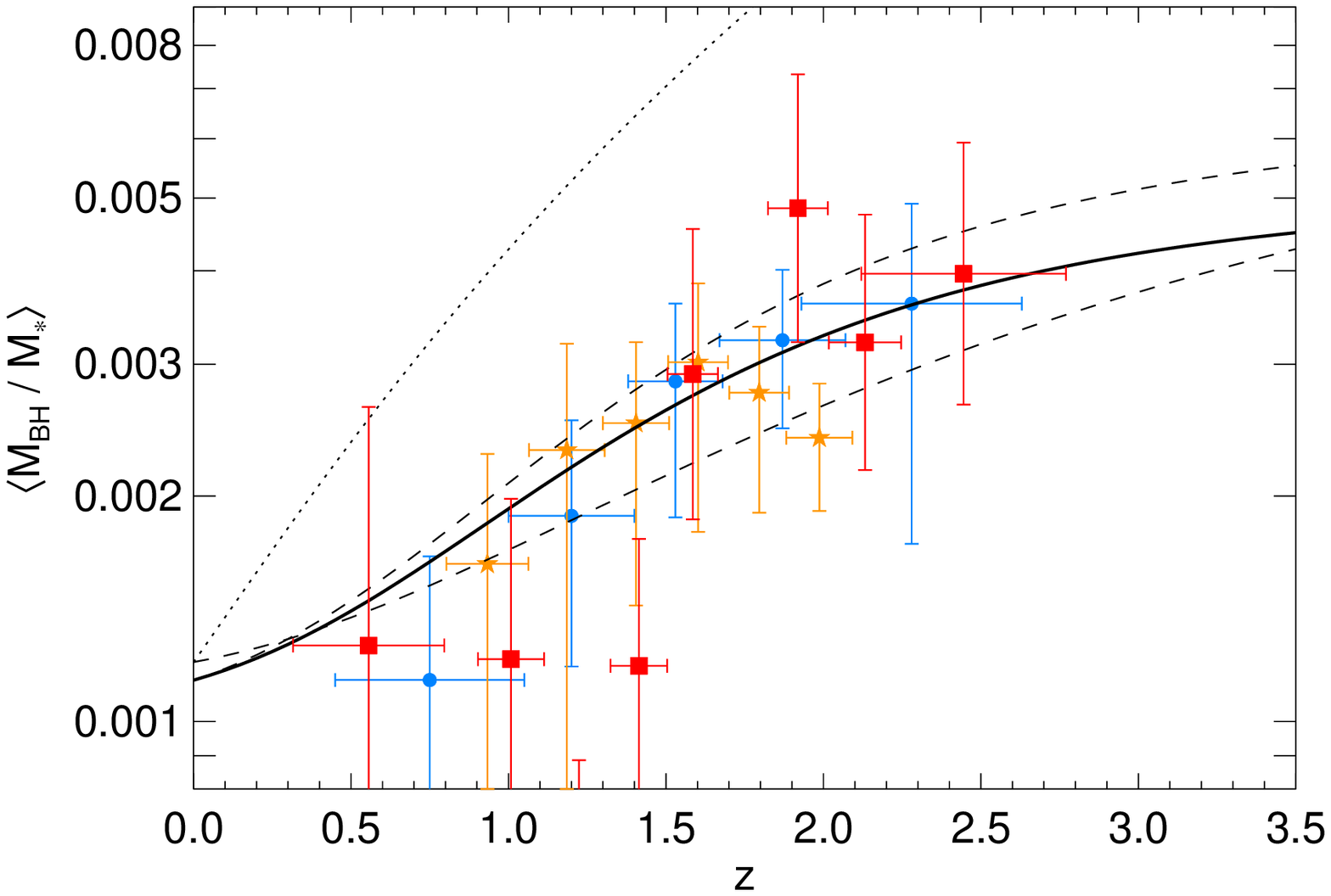}
    \caption{As Figure~\ref{fig:empirical.trujillo} 
    (solid and dashed lines are our prediction and $\pm1\sigma$ range, 
    respectively; dotted line is an alternative $\mbh\propto V_{c}^{4}$ model), 
    but the observationally 
    estimated evolution of $\mbh/\mstar$ (points) is derived from 
    measurements of quasar clustering as a function of redshift, following 
    \citet{fine:mbh.mhost}. For quasars of a 
    given (measured) BH mass, their clustering implies a characteristic host halo mass -- 
    if the ratio of that halo mass to stellar mass does not change with redshift, 
    then the implied $\mbh/\mstar$ evolves as plotted (see \citet{hopkins:clustering} 
    and \citet{fine:mbh.mhost} for 
    details). Measurements of quasar clustering are taken from 
    \citet[][red squares]{fine:mbh.mhost} and 
    \citet[][orange stars]{porciani:clustering.new} from the 2dF, and 
    \citet[][blue circles]{myers:clustering.new} from the SDSS.
    %\citet[][blue circles]{myers:clustering.new} from the SDSS. 
    \label{fig:clustering}}
\end{figure}

We can also attempt to empirically infer the evolution in $\mbh/\mstar$ by considering 
the clustering of quasars as a function of redshift. Essentially, this expands on the 
measurement in \citet{adelbergersteidel:magorrian.evolution}. A more detailed 
discussion of this (and the samples we consider) is given in \citet{hopkins:clustering} 
and \citet{fine:mbh.mhost}, 
but we briefly review it here. 

We consider a quasar sample 
(typically observed near $\sim L_{\ast}$ in the quasar luminosity function) for 
which BH masses or typical Eddington ratios have been directly measured 
(for 2dF quasars, we directly adopt the observed BH mass distributions from 
\citet{fine:mbh.mhost}, otherwise we use those determined as a function of 
quasar luminosity and redshift in \citet{kollmeier:mdot}). We then 
use the observed clustering properties of that sample to infer a characteristic host 
halo mass -- in other words, 
match the observed large-scale bias of the quasar population to the 
average (number-density weighted) large-scale bias for halos in some mass 
range (at the same redshift). We calculate the expected large scale bias of halos 
as a function of mass
following \citet{mowhite:bias} with the improved fitting formulae from 
\citet{shethtormen}. If we assume the ratio of galaxy to halo mass for 
$\sim L_{\ast}$ galaxies does not evolve much with redshift 
\citep[which appears to be 
observationally confirmed to at least $z\sim1$; see e.g.][]{heymans:highz.baryon.fractions,
conroy:z1.hod,zheng:z1.hod}, 
then we have (implicitly) obtained the average host galaxy mass of these quasars, 
for which we know $M_{\rm BH}$ -- i.e.\ an estimate of the mean $\mbh/\mstar(z)$.
Of course, the bias as a function of halo mass 
will depend on the cosmology adopted (specifically the value of $\sigma_{8}$); 
this and e.g.\ the BH mass measurements from quasar spectral energy
distributions
all introduce fairly large systematic uncertainties (at least factor $\sim2$) in the 
absolute implied value of $\mbh/\mstar$. However, to the extent that we are interested only
in the {\em relative} evolution of this with redshift, these uncertainties are much smaller. 

Compiling a number of measurements of quasar clustering as a function of 
redshift, the inferred evolution in $\mbh/\mstar$ is shown in Figure~\ref{fig:clustering}.
This provides a completely independent measurement of 
$\mbh/\mstar$ from that of \citet{peng:magorrian.evolution}, 
with entirely different systematics, but nevertheless is 
in reasonable agreement with their estimates, and with our simple expectation from the 
BHFP relation. Of course, the present observations are not sufficiently 
robust to distinguish between 
evolution by a factor of $\sim2$ versus evolution by a factor of $\sim3$ at 
high redshifts, but different probes seem to suggest a roughly comparable effect to that 
predicted. 

\subsection{A Dissipation-Driven Explanation}
\label{sec:evolution:apriori}

Given the BHFP, we have the empirical expectation that, at high redshift as 
spheroids become more concentrated, $\mbh$ must be larger at fixed $\mstar$. However, 
this does not explain what physically drives these trends. 
In \S~\ref{sec:driving}, we showed that increasing the gas fractions of merger 
progenitors has both of these effects: namely, that by increasing the amount of 
dissipation, more centrally concentrated remnants with 
smaller $\re$, higher $\sigma$ and larger $\mbh$ at fixed $\mstar$
are produced. 
It is both expected and observed that high-redshift disks are characteristically 
more gas-rich, as star formation has simply had less time to operate -- this 
provides a potential {\em a priori} physical motivation for the 
evolution we saw in \S~\ref{sec:evolution:empirical}. 

We therefore briefly examine how systematic evolution in the gas fractions of the 
progenitors of gas-rich, spheroid forming mergers (i.e.\ disk galaxies) should 
affect the resulting $\mbh$-host galaxy correlations of the merger remnants 
(by, for example, increasing the amount of dissipation and therefore forming more 
compact, higher-$\sigma$ remnants). 
To begin, we need to construct an estimate for how the gas fractions of 
typical disks evolve as a function of redshift. Fortunately, the structural properties 
(i.e.\ basic kinematics and the Tully-Fisher relation) of 
disks appear to evolve relatively weakly with redshift, so we can at least attempt to 
simply scale up 
from the properties of local disks (using e.g.\ their star-formation histories). 
Traditionally, the star formation histories 
of local disks are fitted to $\tau$-models, of the form $\mdotstar\propto\exp{[-(t-t_{i})/\tau]}$, 
where $\tau$ is some characteristic timescale and $t_{i}$ is an initial time of formation. 
This is, of course, a non-unique parameterization of the star formation history, but 
nevertheless appears to be a reasonable description of average stellar 
populations \citep[e.g.,][]{noeske:sfh}. 
For a system with stellar mass $\mstar(z=0)=\mstar(t=\tH)$ ($\tH$ being the Hubble time at $z=0$), 
this implies a normalization 
\begin{equation}
\mdotstar(t)=\frac{\mstar(z=0)}{\tau}\,{\bigl(}1-\exp{[(t_{i}-\tH)/\tau]}{\bigr)}^{-1}\,\exp{[(t_{i}-t)/\tau]}.
\end{equation}
It is also well-established that 
disks obey a Kennicutt-Schmidt star formation law \citep{kennicutt98} 
relating the surface density of star formation to 
the gas surface density as $\Sigma_{\rm SF}\sim\Sigma_{\rm gas}^{1.4}$. 
The total SFR $\mdotstar\propto\Sigma_{\rm SF}\,R_{d}^{2}$, and 
$\Sigma_{\rm gas}\propto M_{\rm gas}\,R_{d}^{-2}$, 
where $M_{\rm gas}=\fgas\,M_{\rm tot}=\fgas\,(1-\fgas)^{-1}\,\mstar$ 
and $R_{d}$ is the disk scale length. 
Knowing, therefore, how the mean star formation rate in disks of a given mass evolves, 
and how this scales with the surface density of gas, we need only some estimate of 
their characteristic sizes ($R_{d}$) to infer the (mean) evolution in the total gas mass content. 
If the baryonic Tully-Fisher relation does not evolve with redshift, 
as suggested by observations \citep{conselice:tf.evolution,flores:tf.evolution} -- i.e.\ the 
kinematic structure of disks does not strongly evolve, then 
we have $M_{\rm tot} \propto v_{\rm disk}^{4}$ \citep[][$v_{\rm disk}$ being the 
disk circular velocity $=\sqrt{G\,M_{\rm tot}/R_{d}}$]{belldejong:tf,mcgaugh:tf} and therefore 
$M_{\rm tot}\propto R_{d}^{2}$. Combining 
the Tully-Fisher relation with the Schmidt-Kennicutt law then implies 
\begin{equation}
\frac{\mdotstar(t)}{\mstar(t)}\propto\frac{\fgas^{1.4}}{1-\fgas}.
\label{eqn:ssfr}
\end{equation}
If we demand that the normalization of the Schmidt-Kennicutt law agree 
with the normalization from the fitted $\tau$ model, we then 
arrive at an equation for the implied evolution in $\fgas$, 
\begin{equation}
\frac{\fgas^{1.4}}{1-\fgas}=
%\frac{\fgas(z=0)^{1.4}}{1-\fgas(z=0)}\,
\frac{f_{0}^{1.4}}{1-f_{0}}\,
{\Bigl(}\frac{1-\exp{[(t_{i}-\tH)/\tau]}}{1-\exp{[(t_{i}-t)/\tau]}}{\Bigr)}\,\exp{[(\tH-t)/\tau]}, 
\end{equation}
where $f_{0}=\fgas(z=0)$. 
Adopting the measured best-fit $\tau$ and $z=0$ 
gas fraction $f_{0}$ as a function of 
$z=0$ stellar mass $\mstar$ from \citet{belldejong:disk.sfh} and 
\citet{kannappan:gfs}, respectively, we finally obtain an expected typical $\fgas$ 
and SFR as a function of disk stellar mass at any cosmic time $t$. 

\tableclear
\begin{figure}
    \centering
    \figexpand
    \plotter{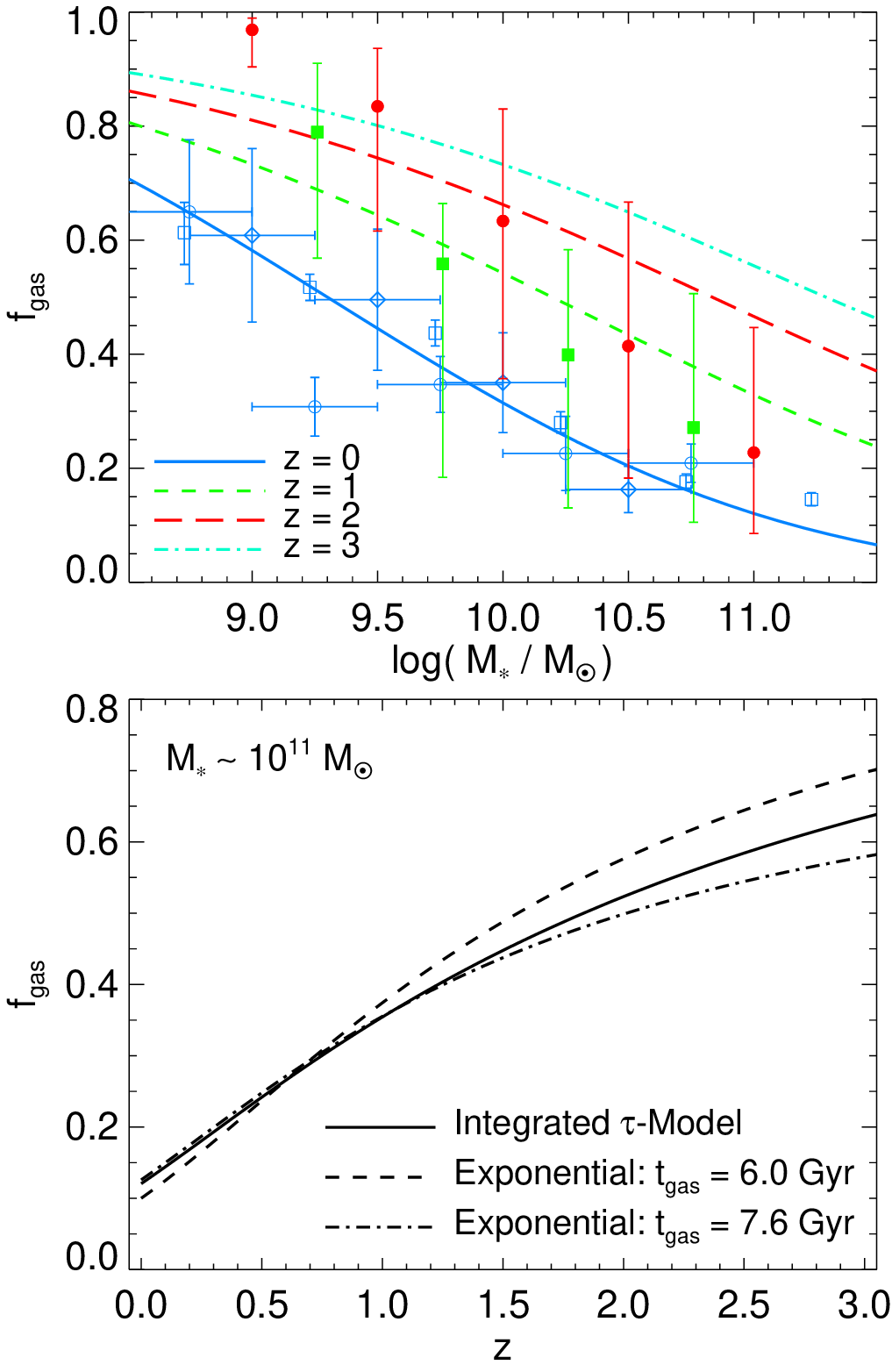}
    \caption{{\em Upper:} Comparing the predicted evolution in characteristic disk gas 
    fractions (as a function of stellar mass) from our simple toy model (in which 
    we integrate backwards in time the best-fit 
    $\tau$-model star formation histories of local disks) with that observed. 
    We compare the expected mean $\fgas(\mstar)$ with 
    estimates at $z=0$ 
    \citep[blue points;][as diamonds, squares, \&\ , circles respectively]
    {belldejong:tf,kannappan:gfs,mcgaugh:tf}, 
    $z\sim1$ \citep[green squares; estimated from H$\alpha$ luminosities][]{shapley.z1.abundances}, 
    and $z\sim2$ \citep[red circles][]{erb:lbg.gasmasses}. 
    {\em Lower:}
    Evolution in $\fgas$ from our simple empirical model for a 
    constant $\mstar=10^{11}\,M_{\sun}$, compared to a simple exponential 
    history with $e$-folding time $t_{\rm gas}$ (as labeled).
    \label{fig:fgas.evolution}}
\end{figure}

Figure~\ref{fig:fgas.evolution} compares the expected 
 $\fgas(\mstar(t))$ from this 
parameterization with that observed at a number of different redshifts. 
We also compare 
this estimate to another, even simpler parameterization we could adopt: assuming an 
isolated disk obeying a $\tau$-model, with $\fgas=1$ at $t=0$ and $\fgas\approx0.1$ 
(appropriate for a Milky Way-like $\sim L_{\ast}$ disk today) at $z=0$. 
This implies an exponential growth of $\fgas$ with lookback time, with $e$-folding 
time $\sim 6\,$Gyr. 
The results are similar, and appear to describe the observed 
$\fgas$ evolution reasonably well. We have also checked that the expectation 
from Equation~(\ref{eqn:ssfr}) is consistent with observed specific 
star formation rates as a function of $\mstar$ from $z=0-3$ 
\citep[see, e.g.][]{bauer:ssfr,feulner:ssfr,papovich:ssfr}, and find good agreement
even up to $z\sim3$ (which should 
not be surprising, as this essentially just says that the $\tau$ model is 
indeed a reasonable description of the mean star formation history). 

\begin{figure}
    \centering
    \figexpand
    \plotter{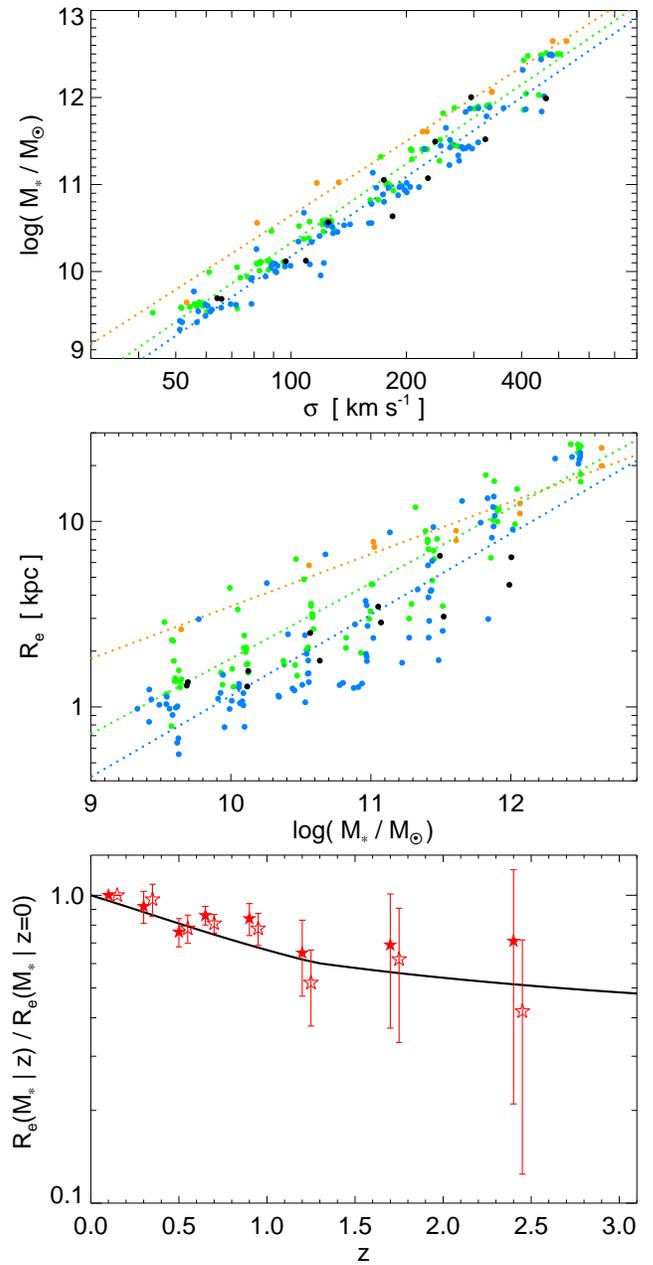}
    \caption{{\em Upper:} Faber-Jackson ($\mstar-\sigma$) and 
    $\re-\mstar$ relations in our simulations as a function of gas fraction 
    ($\fgas=0.2,\ 0.4,\ 0.8,\ 1.0$ in orange, green, blue, and black points, 
    respectively, with the best-fit trend to each as the dotted line of 
    corresponding color). At fixed $\mstar$, higher-$\fgas$ systems have 
    smaller $\re$ and larger $\sigma$, as in Figure~\ref{fig:fgas.casestudy}. 
    {\em Lower:} Given our empirical estimate of the evolution in $\fgas$ 
    at fixed $\mstar=10^{11}\,M_{\sun}$, and the dependence of $\re(\mstar)$ 
    on $\fgas$ (upper panel \& Figure~\ref{fig:fgas.casestudy}), solid 
    line plots the expected evolution in $\re(\mstar)$ with redshift. 
    Points compare the observed evolution from 
    \citet{trujillo:size.evolution}, for $\mstar>3\times10^{10}\,M_{\sun}$ (filled) and 
    $\mstar>6\times10^{10}\,M_{\sun}$ (open). 
    The increasing gas content in high-redshift merger progenitors 
    predicts their size evolution at fixed stellar mass. 
    \label{fig:fj.evolution}}
\end{figure}

Given this evolution in $\fgas$, we are now in a position to estimate 
how this should change the effective radii and velocity dispersions of 
merger remnants, and (given the BHFP) the average value of 
$\mbh$ at fixed stellar mass $\mstar$. Figure~\ref{fig:fj.evolution} 
shows the evolution of the $\mstar-\sigma$ and $\re-\mstar$ relations 
with $\fgas$, from our simulations, where the behavior is similar 
to that we discussed in \S~\ref{sec:driving}. Combining the trend in 
$\re(\mstar)$ as a function of $\fgas$ from our simulations, and the trend in 
$\fgas(z)$ above, we also then predict how $\re(z)$ should 
evolve, at fixed $\mstar$. Comparing this to the results from \citet{trujillo:size.evolution} shows 
excellent agreement given our simple estimate of $\fgas(z)$. 

Likewise, either directly adopting our fitted trend of $\mbh/\mstar$ as a 
function of $\fgas$ from Figure~\ref{fig:residuals.fgas}, or using our 
estimate of how $\re$ (and, correspondingly, $\sigma$) evolve 
at fixed $\mstar$ with $\fgas$ and applying them to the BHFP relation 
$\mbh\propto \mstar^{0.7}\,\sigma^{1.5}\propto \mstar^{1.5}\,\re^{-1}$, 
this predicts that the mean $\mbh/\mstar$ should increase with redshift. 
Figures~\ref{fig:empirical.trujillo} \&\ \ref{fig:clustering} show the evolution in $\mbh/\mstar$ expected 
from this simple derivation. Again, the agreement with the 
observationally estimated rate of evolution in $\mbh/\mstar$ is 
good. 

We do caution that our estimate of the mean evolution in $\fgas$, while consistent with the 
observational constraints on gas fractions and specific star formation 
rates at all redshifts with which we compare ($z\lesssim3$), is only intended as a 
rough lowest-order approximation to realistic disk evolution. It is of course 
possible that other disk properties, such as the Tully-Fisher relation, 
begin to evolve at high redshift \citep[see e.g.][]{genzel:highz.disk}, 
or that elliptical formation at high redshift might proceed 
in more chaotic multiple mergers for which our simple approximations 
are not valid, especially in the most massive systems at very high redshifts $z\sim6$. 
However, we have seen in Figure~\ref{fig:residuals.initialconditions} 
that evolution in properties such as virial velocities, disk structure, and 
orbital parameters do not drive much additional evolution in $\mbh/\mstar$. 
Furthermore, some reassurance comes from 
\citet{li:z6.quasar}, who consider simulations which 
adopt cosmologically-derived merger histories for the brightest 
$z\sim6$ quasars, involving multiple major and minor mergers, and find that the 
remnant obeys a similar $\mbh/\mstar$ relation to our idealized high-$\fgas$ simulations 
-- i.e.\ it is consistent with our estimated evolution in $\mbh/\mstar$ 
(and $\re/\mstar$) as a function of 
redshift. Despite a chaotic, rapid merger history, the remnant properties can 
approximately be predicted as a function of stellar mass and gas fraction 
from our simple prescription. The most important (at lowest order) 
element of our estimate is simply the qualitative statement that the progenitors 
of high-redshift ellipticals should be characteristically more gas-rich -- 
which given the rapid star-formation timescales estimated for massive 
disks \citep{belldejong:disk.sfh} and observationally inferred brief, potentially burst-dominated 
star-formation histories of the most massive ellipticals \citep[e.g.,][]{thomas}, 
is naturally expected in theories of hierarchical growth. In other words, so long as the 
total stellar mass being violently scattered and total gas supply reaching the spheroid 
center (i.e.\ total mass and gas fraction) are similar, the details of the merger 
history and distinction between binary or multiple mergers should not 
significantly change the simple gravitational physics which determine the 
most basic elements of the remnant structure.
We therefore suggest that evolution in $\fgas$ is indeed the dominant (although not the only) 
physical agent driving evolution in the $\mbh-\mstar$ relation.

\section{Discussion}
\label{sec:discussion}

Using a large set of numerical simulations of major galaxy-galaxy mergers, 
which include the effects of gas dissipation, 
cooling, star formation, and black hole accretion, we 
find that a feedback-driven model of BH growth and self-regulation
predicts the existence of a BH ``fundamental plane'' (BHFP), 
of the form $\mbh\propto\sigma^{3.0}\,\re^{0.5}$ or 
$\mbh\propto \mstar^{0.5-0.7}\,\sigma^{1.5-2.0}$, 
analogous to the FP of spheroids. 
Comparing with existing BH 
mass measurements, the observed systems appear to follow a nearly identical BHFP relation. 
Specifically, there are significant (at $>99.9\%$ confidence) trends in the residuals of 
the $\mbh-\sigma$ relation with $\mstar$ and $\re$ at fixed $\sigma$, 
and likewise in the $\mbh-\mstar$ relation (with $\sigma$ or $\re$ at fixed $\mstar$). 
While changes in halo circular velocity, merger orbital parameters, progenitor disk 
redshifts and gas fractions, ISM gas pressurization, and other parameters can drive 
changes in e.g.\ $\sigma$ at fixed $\mstar$, and therefore change in the $\mbh-\sigma$
or $\mbh-\mstar$ relations, the BHFP is preserved.

This provides a new paradigm for understanding the traditional relations between 
BH mass and either bulge velocity dispersion or mass. These correlations (as 
well as those with other bulge properties such as effective radius, 
central potential, dynamical mass, concentration, Sersic index, and bulge binding 
energy) are all projections of the same fundamental plane relation. 
Just as the Faber-Jackson relation between e.g.\ stellar mass or luminosity 
and velocity dispersion ($\mstar-\sigma$) is understood as a projection of the 
more fundamental relation between $\mstar$, $\sigma$, and $\re$, 
so too is the $\mbh-\sigma$ relation ($\mbh\propto\sigma^{4}$) a projection of the 
more fundamental relation $\mbh\propto \sigma^{3}\,\re^{0.5}$. Recognizing this 
resolves the nature of several apparent outliers in the $\mbh-\sigma$ relation, 
which simply have unusual $\sigma$ values for their stellar masses 
or effective radii, and eliminates the strong correlations between residuals 
(in both observations and simulations). While the various changes above in 
merger properties 
can and do bias the various projections of the BHFP to different values, they simply move 
remnants {\em along} the BHFP relation.

Given the empirical tendency towards more compact, smaller-$\re$ spheroids 
at fixed stellar mass $\mstar$ at high redshift, the BHFP predicts that BHs should be 
more massive at fixed $\mstar$. \citet{trujillo:size.evolution} 
compile a number of observations of the sizes of early-type galaxies at fixed 
stellar mass (for typical $\sim L_{\ast}$ galaxies), and find a best-fit trend 
$\re\propto (1+z)^{-0.45}$. The observed BHFP predicts that BH mass 
scales roughly as $\propto\mstar^{1.5}\,\re^{-1.0}$, which yields the 
prediction that the typical hosted BH mass at fixed 
stellar mass (or ratio of $\mbh/\mstar$) should increase as $(1+z)^{0.5}$. This agrees 
well with recent direct estimates of the BH to host stellar mass ratio 
at high redshift \citep{peng:magorrian.evolution}, as well as indirect estimates of 
the evolution in the mean $\mbh/\mstar$ from 
comparisons of quasar luminosity functions and early-type mass density 
measurements \citep{merloni:magorrian.evolution}, BH and spheroid 
mass functions \citep{hopkins:msigma.limit}, 
and quasar clustering as a function of redshift \citep{adelbergersteidel:magorrian.evolution,
wyithe:magorrian.clustering,hopkins:clustering,fine:mbh.mhost}. 
Interestingly, if we consider this in greater detail, 
observations suggest that the evolution in spheroid sizes is relatively weak to 
$z\sim0.8$ \citep{mcintosh:size.evolution} and stronger from $z\sim1-2$. Our BHFP 
analysis argues that the same should be true for the ratio of BH mass to stellar mass, 
and indeed \citet{peng:magorrian.evolution} note that there is no 
significant evolution at lower redshifts $z\lesssim1$ in their sample, compared 
to the substantial evolution they observe at $z\sim1-3$. 

We have also developed a physically motivated model for this evolution. Based on the 
empirical and theoretical expectation 
that the progenitor disks in typical mergers should be more gas-rich at 
higher redshifts, we expect mergers to be more dissipational, yielding more concentrated 
remnants and driving the evolution in $\mbh/\mstar$ along the BHFP. Indeed, adopting an 
empirical estimate for the mean $\fgas$ as a function of stellar mass and redshift, 
we predict a trend with redshift in the size-mass relation of our merger remnants 
which is similar to the observations compiled in \citet{trujillo:size.evolution}, 
and consequently a trend in $\mbh/\mstar$ like that observed by \citet{peng:magorrian.evolution}.
Our simulations thus provide critical support to arguments
from a semi-analytic context, such as those made by  
\citet{khochfar:size.evolution.model}, that the observed evolution in $\re(\mstar)$ can be 
explained by the increasingly gas-rich, dissipational nature of merger progenitors at high redshifts. 
It is worth noting that, although it does not rule out such mergers occurring, the trend 
in $\re(\mstar)$ can be explained entirely by changing gas fractions in gas-rich, dissipational 
mergers, {\em without} invoking subsequent ``dry'' mergers at low redshifts to increase $\re$
(see also, Krause et al, 2007, in prepartion). 

We also emphasize that our results are entirely consistent with the previous study of 
\citet{robertson:msigma.evolution}. However, in that case, the authors considered only the effects of 
the (relatively weak) scaling of disk sizes at fixed $\mstar$ with redshift, and found 
that this introduced a weak evolution in the $\mbh-\sigma$ relation. Allowing for $\fgas$ to 
scale systematically with redshift drives the 
evolution in $\mbh/\mstar$ which is analyzed herein, and placing both simulations and observations 
in the context of the FP relation reconciles the apparent disagreement 
between the predictions of \citet{robertson:msigma.evolution} for the $\mbh-\sigma$ relation 
and observations by e.g.\ \citet{peng:magorrian.evolution} of the high-redshift $\mbh-\mstar$ relation. 

There are a number of direct, testable predictions of this fundamental plane 
model for the correlations between BH and host properties. At both low and 
high redshifts, systems should lie on the same BHFP. Therefore, measurements of 
the effective radii or velocity dispersions of the \citet{peng:magorrian.evolution} objects 
should find that they are more compact (smaller $\re$, larger $\sigma$) than their 
$z=0$ counterparts of the same stellar mass, in a manner consistent with 
the BHFP in Table~\ref{tbl:correlations}. If it is really the BHFP driving the apparent 
evolution in $\mbh/\mstar$ with redshift, i.e.\ the fact that at fixed $\mstar$, higher-redshift 
systems are more compact, then this also predicts different evolution 
for BH mass relative to $\sigma$ (the $\mbh-\sigma$ relation) than for BH mass 
relative to $\mstar$. Adopting the \citet{trujillo:size.evolution} estimate for 
how $\re$ scales with redshift and the near-IR spheroid fundamental plane of 
\citet{pahre:nir.fp} to relate $\sigma$ and $\re$ at fixed $\mstar$, along with our 
BHFP in terms of $\mstar$ and $\sigma$, this predicts a trend of the form 
$\mbh/\sigma^{4}\propto(1+z)^{-0.25}$, i.e.\ weaker and {\em inverse} evolution in 
$\mbh/\sigma$ at fixed stellar mass, quite similar to the predictions made 
by \citet{robertson:msigma.evolution} and consistent with the 
observations of \citet{shields03:msigma.evolution}. 

At low redshifts, improved measurements of the host properties of systems with 
well-measured BHs can significantly improve constraints on the BHFP. As noted 
in Table~\ref{tbl:correlations}, the present observations demand a correlation of the 
form $\mbh\propto\sigma^{\alpha}\,\mstar^{\beta}$ over a simple correlation with 
either $\sigma$ or $\mstar$ at $\gtrsim3\,\sigma$ confidence. Already, 
this puts strong constraints on theoretical models of BH growth and evolution -- 
BH mass does not simply scale with the star formation (stellar mass) 
or virial velocity of the host galaxy. 
However, there is still a 
substantial degeneracy between the slopes $\alpha$ and $\beta$ (roughly 
along the axis $\beta\approx1-\alpha/4$). For example, the current data do not 
allow us to significantly distinguish a pure correlation with spheroid binding 
energy $\mbh\propto (\mstar\,\sigma^{2})^{2/3}$ from the marginally favored 
relation $\propto \mstar^{1/2}\,\sigma^{2}$ -- both suggest 
that the ability of BHs to self-regulate 
their growth must be sensitive to the potential well at the 
center of the galaxy (and therefore to galactic structure), but the difference 
could reveal variations in the means by which BH feedback couples 
to the gas on these scales. 

Increasing the observed sample sizes and, 
in particular, extending the observed baselines in mass and $\sigma$ will 
substantially improve the lever arm on these correlations. In particular, the addition 
of stellar mass $\mstar$ information to the significant number of objects which have 
measurements of $\sigma$ and indirect measurements of $\mbh$ from 
reverberation mapping would enable considerably stronger tests of our 
proposed BHFP relation. Furthermore, to the extent that the evolution in the 
$\mbh/\mstar$ and $\re(\mstar)$ relations is driven by the relatively gas-rich 
nature of the merger progenitors, the residuals in e.g.\ $\mbh/\mstar$ should also 
be correlated with other tracers of the amount of dissipation in the spheroid-forming 
merger. These include quantities such as $\re$ and $\sigma$, of course, in the 
context we have discussed, but also e.g.\ the central phase-space density, 
kinematic properties such as rotation and kinematically decoupled components 
\citep[see][]{hernquist:kinematics,cox:kinematics}, and potentially the presence of central cusps in the 
stellar light profiles of the remnants \citep{mihos:cusps}. 
We do note the caveat from \S~\ref{sec:driving}, however, 
that care should 
still be taken to consider only bulge properties and remove e.g.\ rotationally supported 
contributions to the velocity dispersion. 

Given the robust nature of the BHFP, we might also ask if there are processes 
which we might expect to drive systems away from the BHFP.  For example, what are 
the effects of subsequent gas-poor (spheroid-spheroid or ``dry'') mergers on the BHFP? 
Such mergers, by definition, conserve total stellar mass and BH mass (simply 
adding the $\mstar$ and $\mbh$ of the two merged systems). However, simple 
energetic arguments imply that $\sigma$ is not dramatically changed
\citep[e.g.,][]{hernquist:phasespace,nipoti:dry.mergers,boylankolchin:dry.mergers}. 
If we assume $\sigma$ is unchanged in a ``dry'' merger, then the BHFP relations 
in Table~\ref{tbl:correlations} imply a $\sim0.08-0.12$\,dex offset (in the sense of 
$\mbh$ being too large) from a major (equal-mass) dry merger (the offset being 
$\sim0.03-0.05$\,dex for a more probable 1:3 mass-ratio merger). This is also 
supported by a small subset of full numerical re-merger simulations from 
\citet{robertson:fp}. Although this 
implies that the BHFP will not survive a large number ($\gtrsim3$) 
of successive dry mergers, the observationally 
estimated rate of $\sim0.5-1$ major dry mergers for a typical massive elliptical 
\citep{bell:dry.mergers,vandokkum:dry.mergers} implies that the realistic resulting deviations from 
the BHFP are small (smaller than the internal scatter in the relation itself). 
This may be important, however, for explaining how the compact, high-redshift spheroids 
observed and predicted herein increase in size to lie on the local $\re-\mstar$ relations 
(since each major dry merger will approximately double $\re$), a possibility 
that will be investigated in detail in future work.

Therefore, as appears to be borne out by the local observations we consider, the 
BHFP appears to be a robust correlation, which provides an improved 
context in which to understand the nature and evolution of the numerous observed 
correlations between BH and host spheroid properties.  In particular, the results
described here provide new, important 
constraints for models of BH growth, feedback and self-regulation, and  
support the proposal developed by \citet{hopkins:qso.all} that
major mergers between gas-rich galaxies represent the principle mechanism for
triggering intense starbursts \citep[e.g.,][]{barnes:starbursts,mihos:starbursts.94,mihos:starbursts.96}
that evolve into quasars \citep[e.g.,][]{sanders:review} and which eventually
leave remnants satisfying the same structural correlations observed for 
elliptical galaxies \citep[e.g.,][]{robertson:fp}.

\acknowledgments We thank Chien Peng for illuminating discussion and comments. 
This work was supported in part by NSF grant AST
03-07690, and NASA ATP grants NAG5-12140, NAG5-13292, and NAG5-13381.

\bibliography{ms}

\end{document}